\documentclass{aa}
\usepackage{graphicx,float} 
\usepackage{txfonts} 
\usepackage{rotating} 
\def\OmM{\Omega_{\rm M}}
\def\OmL{\Omega_{\Lambda}}

\def \nh{N$_{H}$~} 
\def \ergcms {erg~cm$^{-2}$~s$^{-1}$ } 
\def \ergs {erg~s$^{-1}$ } 
\def \LxTx {$L_{\rm x}-T_{\rm x}$ }
\begin{document} 

%??I made another iteration on the title.

\title{ The XMM--NEWTON $\Omega$ Project: I. The X-ray Luminosity --
Temperature Relation at $z>0.4$}
 
\author{D. H. Lumb \inst{1} \and J.G. Bartlett \inst{2,3} \and A.K. Romer \inst{4}  \and  
A. Blanchard \inst{5} \and D.J. Burke \inst{6} \and C.A. Collins 
\inst{7} \and R.C. Nichol \inst{4} \and   \\
M. Giard \inst{8} \and P.B. Marty \inst{5} \and J.  
Nevalainen \inst{6} \and R. Sadat \inst{5} \and S. C. Vauclair \inst{5}} 
 
\institute{Science Payloads and Advanced Concepts  Office, European Space Agency, ESTEC, 2200AG Noordwijk, The Netherlands 
\and 
APC - Universit\'e Paris 7/PCC - Coll\`ege de France, 11, pl. Marcelin Berthelot
        F-75231 Paris Cedex 05, France
\and
        Centre de Donn\'ees astronomiques de Strasbourg, 11, rue de 
        l'Universit\'e, F-67000 Strasbourg, France
\and 
Physics Department, Carnegie Mellon University, Pittsburgh, PA 15213,
USA 
\and 
Laboratoire d'Astrophysique, OMP, CNRS, UPS, 14, Av Ed. Belin, 31 400, Toulouse, France 
\and 
Harvard-Smithsonian Center for Astrophysics, 60 Garden Street, Cambridge, MA 02138. USA 
\and 
Astrophysics Research Institute, Liverpool John Moores University, Twelve Quays House, Egerton Wharf, Birkenhead CH41 1LD, UK 
\and 
Centre d'Etude Spatiale des Rayonnements, 9 avenue du Colonel Roche, BP 4346, 31028 Toulouse, France  }

\offprints{D Lumb (dlumb@rssd.esa.int)} 
\date{Received date / Accepted date} 
\titlerunning{Observations of high-z clusters}

%?? add the radius to which the spatial fit was done to each cluster section 

%?? once Table 5 is finalized, search for ``row'' to check that the row numbering
% in the text is correct e.e. in section 3.5

%?? Gamma used in Figure 1 and A more common for L-T evolution in other papers
%  so consider change all the Gamma's to A

%??I reworked the abstract, but not very carefully

\abstract{We describe XMM-Newton Guaranteed Time observations of a
sample of eight high redshift ({\it 0.45$<$z$<$0.62}) clusters. The goal of
these observations was to measure the luminosity and the temperature of
the clusters to a precision of $\sim$10\%, leading to constraints on the
possible evolution of the luminosity--temperature ($L_{\rm x}-T_{\rm
x}$) relation, and ultimately on the values of the matter density,
$\OmM$, and, to a lesser extent, the cosmological constant $\OmL$. The
clusters were drawn from the SHARC and 160 Square Degree (160SD) ROSAT
surveys and span a bolometric (0.0--20~keV) luminosity range of 2.0 to
14.4 $\times 10^{44}$ \ergs (H$_{o}$=50,$\OmM$=1, $\OmL$=0).  Here we describe our data analysis
techniques and present, for the first time with XMM-Newton, a $L_{\rm x}-T_{\rm x}$ relation.  For each of
the eight clusters in the sample, we have measured total
($r<r_{virial}$) bolometric luminosities, performed $\beta$-model fits
to the radial surface profiles and made spectral fits to a single
temperature isothermal model. We describe data analysis techniques
that pay particular attention to background mitigation. We have also estimated temperatures and
luminosities for two known clusters (Abell 2246 and RXJ1325.0-3814),
and one new high redshift cluster candidate (XMMU J084701.8+345117),
that were detected off-axis. Characterizing the $L_{\rm x}-T_{\rm x}$
relation as $L_{\rm x} = L_{6} (\frac{T}{6 keV})^{\alpha}$,
%(1+z)^{A}, 
we find $L_{6}=15.9 ^{+7.6}_{-5.2} \times
10^{44}$\ergs~ and $\alpha$=2.7$\pm$0.4 for an $\OmL=0.0, \OmM=1.0$,
$H_0=50$ ~km~s$^{-1}$~Mpc$^{-1}$ cosmology 
at a typical redshift $z \sim 0.55$.  
Comparing with the low
redshift study by \cite{markevitch}, we find $\alpha$ to be in agreement, and assuming $L_{\rm x}-T_{\rm x}$
to evolve as $(1+z)^{A }$, we find $A$=0.68$\pm$0.26
for the same cosmology and $A=1.52^{+0.26}_{-0.27}$ for an
$\OmL=0.7, \OmM=0.3$ cosmology. Our $A $ values are very similar to
those found previously by \cite{Vikh_hiz} using a compilation of
Chandra observations of $0.39<z<1.26$ clusters.  We conclude that
there is now evidence from both XMM-Newton and Chandra for an
evolutionary trend in the $L_{\rm x}-T_{\rm x}$ relation. This
evolution is significantly below the level expected from the
predictions of the self-similar model for an $\OmL=0.0, \OmM=1.0$,
 cosmology, but  consistent with self-similar model in an 
$\OmL=0.7, \OmM=0.3$ cosmology. 
Our observations lend support
to the robustness and completeness of the SHARC and 160SD surveys.

%?? is Gamma=0.65 valid for the L_6=16.4 fit; it wasn't presented 
% simultaneously with it in section 5
%%AB I do not think we can have a different L_6 and (1+z)^{\Gamma}
%%AB in the same formula. 

%?? we should give our L_6 and alpha values and Ho for the Lambda cosmo
%%AB no I think only \Gamma is meaningful. Given that all clusters
%%AB are        at the ~ same redshift we can give the normalization.

}

\maketitle 
 
\section{Introduction}
\subsection{Motivation} 
 As the most massive gravitationally bound objects in the universe,
galaxy clusters are particularly sensitive to the evolution of the
density perturbations responsible for their formation.  Cluster
abundance as a function of mass and redshift is dictated by the mass
function (\cite{ps}; \cite{jenkins}), which gives the comoving space
density of collapsed objects as a function of mass and redshift.  In
standard models, the mass function falls off as a Gaussian at the
high mass end, reflecting the Gaussian nature of the density
perturbations.  Galaxy cluster abundance is therefore exquistely
sensitive to the amplitude of the perturbations and its evolution.
Since this evolution is controlled by the underlying cosmological
background, observations of cluster abundance offer an effective way
to constrain certain cosmological parameters, such as the density
parameter (\cite{ob1};\cite{bb})
or the dark energy parameter and equation--of--state (\cite{Majumdar};\cite{Wang}).  The present day cluster abundance is degenerate in the
matter density and amplitude of the power spectrum; evolution of the
abundance breaks this degeneracy.  Constraints obtained in this manner
are complementary to others that essentially rely on determinations of
cosmological distances; for example, those found by observations of
supernovae type Ia or by measurements of cosmic microwave background
anisotropies.

Cluster mass is, however, difficult to measure directly, and in
practice one seeks a direct observable that is closely related to
virial mass. Lensing surveys would seem the most suited to the task,
as the effects of lensing are of course directly related to mass
(albeit projected along the line--of--sight). However, cluster mass
estimates from weak and strong lensing remain controversial as they
suffer from several systematic uncertainties (e.g. projection effects).  Among X--ray
observables, intracluster gas temperature is expected to be tightly
correlated with virial mass, an expectation borne out by simple
hydrostatic considerations as well as by numerical
simulations~(\cite{evrard}; \cite{bn}). X-ray luminosity, on the other
hand, is a much less robust mass indicator, despite being
significantly easier to measure, because it depends on the density
profile of the intracluster gas, the physics of which is currently
difficult to model.

\begin{figure}
\begin{center}
\resizebox{\hsize}{!}{\includegraphics{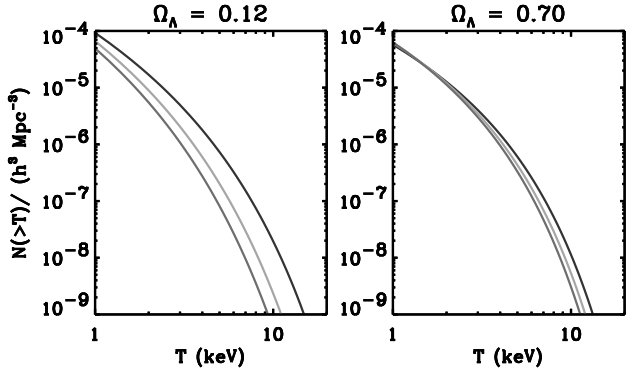}}
\caption{The predicted X--ray temperature function at redshifts
$z=0.05$ (black), 0.33 (light grey) and 0.5 (dark grey) for a
flat high density model (left; $\OmL=0.12$) and a concordance model
(right; $\OmL=0.7$), both fitted to the local ($z \sim 0.05$)
temperature function.}
\label{fig:dndT}
\end{center}
\end{figure}

With a calibrated $T_{\rm x}-M$ (temperature--mass) relation, the mass
function can be translated into an observable cluster temperature
distribution function $dn/dT$.  The exact $T-M$ relation to use is of
course a key ingredient, one that may be addressed, for example, using
numerical simulations, or directly from detailed observations that
determine both cluster mass and temperature (e.g.~\cite{jukka}).
Figure~\ref{fig:dndT} compares temperature function predictions for a
high matter density and a flat model, both normalized to the
present--day, observed $dn/dT$ (e.g. \cite{ha};
\cite{edge}; \cite{b2000}; \cite{ikebe01}).
Evolution toward higher redshift is strikingly different in two
cosmological models {\it differing by their matter content alone} and illustrating
the above argument.

Despite this obvious promise, it has been difficult to use $dn/dT(z)$
to explore the cosmological model because direct measurements of
$T_{\rm x}$ at $z>0$ require long integration with satellite
observations.  Prior to the launch of Chandra and XMM-Newton, the
number of measured cluster temperatures at high redshifts was very
small.  This is attested to by the fact that the most distant
determination of $dn/dT$ (\cite{h00}) to date was based on a sample of
only fourteen $z>0.3$ clusters  (mean $\sim$0.38) from the Einstein Medium Sensitivity
Survey (EMSS;
\cite{emss}). 

An alternative, but related approach is to apply a
luminosity--temperature ($L_{\rm x}-T_{\rm x}$) relation to a
flux--limited sample, thereby obtaining either the temperature
function, or a redshift distribution at given temperature (\cite{ob2};\cite{sbo};\cite{reichart};\cite{borgani99};\cite{borgani01}). The 
advantage of this approach is that it does not require detailed X-ray
spectroscopy of all the clusters in a flux-limited sample to establish
the $L_{\rm x}-T_{\rm x}$ relation and its evolution.  There are now
several flux limited catalogs of medium to high redshift clusters, in
addition to the EMSS, to which one can apply this technique. These
include those based on data from the ROSAT All Sky Survey
(e.g. \cite{MACS}; \cite{Zhang}), NEP (e.g. \cite{nepg}; \cite{neph});
and from the ROSAT pointing archive (e.g. \cite{brightsharc2};
\cite{southern}; \cite{Vikh}; \cite{Mullis03}; \cite{WARPS}; \cite{RDCS}).  

\subsection{The XMM--Newton $\Omega$--Project}
The goal of the XMM--Newton $\Omega$--project (\cite{Bartlett}) is to
increase the number of high quality X-ray cluster temperature
measurements at $z>0.3$, and thus enhance the scientific yield from
the various ROSAT surveys. We describe below XMM-Newton observations
of eight ROSAT clusters at $z>0.4$ (median of $z=0.54$) performed as
part of the Guaranteed Time programme. Related observations conducted
during the open time programmes of seven $0.3<z<0.4$ clusters are
discussed elsewhere (\cite{opentime1}; \cite{opentime2}).
Cosmological interpretation is presented in \cite{XMM2}. In
Sections 2 \& 3 we describe the observations and data analysis
techniques. In Section 4 we discuss each of the eight clusters in
turn. In Section 5 we present our $L_{\rm x}-T_{\rm x}$ relation and
compare it to previous work. 

This paper serves also to introduce the Project, and represents an
opportunity to provide detailed descriptions of analysis techniques
used in the Project and which can be be used generally for XMM-Newton
observations of cluster targets. 
Except where explicitly stated elsewhere we
use a $\OmM=1$, $\OmL=0$, $q_0=0.5$ model with 
$H_0=50$~km~s$^{-1}$~Mpc$^{-1}$ that has been
most frequently used in the past as the parameter set for X-ray
cluster studies. When necessary for examining cosmological
implications we correct our results 
to a concordance model.

\section{Observation Programme} 
\subsection{The Sample}
Table~\ref{tab:observns} summarizes the locations, date of
observations and other details of the eight clusters in this
program. Seven of the eight clusters were drawn from the SHARC
Surveys, four from the Southern SHARC (\cite{southern}) sample and
three from the Bright SHARC (\cite{brightsharc2}) sample.  The
SHARC cluster samples are based on searches for clusters
serendipitously detected in ROSAT PSPC observations. They complement
each other in that they cover, respectively, 17.7 degrees$^{2}$ to a
flux limit of $\simeq 3.9\times 10^{-14}$ erg~s$^{-1}$cm$^{-2}$
(\cite{deepsharc2}) and 178.6 degrees$^{2}$ to a flux limit of $\simeq
1.4\times 10^{-13}$ erg~s$^{-1}$cm$^{-2}$ (\cite{brightsharc1}).  This
strategy has yielded a combined cluster catalogue that straddles $L^*$
over the redshift range $0.2 < z < 0.8$ and shows a consistent picture
of a non-evolving cluster luminosity function (\cite{deepsharc2};
\cite{deepsharc1}) except, possibly, at luminosities greater than
$L_{x}=5\times 10^{44}$ erg~s$^{-1}$ (\cite{brightsharc1};
\cite{adami}). Two aspects of the SHARC surveys, makes them
particularly well suited to the XMM-Newton $\Omega$ project; they have
been subjected both to detailed optical follow-up
(\cite{brightsharc2}; \cite{southern}) and to extensive selection function
simulations (\cite{adami}; \cite{southern}). The eighth cluster in our
observation programme was taken from the 160 Square Degree ROSAT
Survey (160SD hereafter, \cite{Vikh} and \cite{Mullis03}). There is
considerable overlap, both in terms of methodology and cluster
members, between the 160SD and the SHARC surveys (3 of the 7 SHARC
clusters are also members of the 160SD catalogue), however, this
particular cluster (RXJ0847.2) was not a member of either SHARC sample
because its host PSPC observation did not meet the SHARC exposure time
criterion.  The selection of clusters sampled in this programme was driven
solely  by a requirement  to observe all z$\geq$0.5 targets from the SHARC
  surveys. Following the   visibility and observability constraints of
  the XMM-Newton Guaranteed   Time programme we added two clusters
  with slightly lower redshift from the SHARC sample in addition to
  the one from the 160SD catalogue.
 
\small 
\begin{table*}
\begin{center} 
\begin{tabular}{lcl l l l l l l l}\hline\hline 
Cluster ID&RA&Dec&z&Date&Obs ID&Duration&Filter\\ \hline 
RXJ0337.7--2522 &03:37:45&-25:22:26&0.577$^{s}$&2001-08-18T11:46:53&0107860401&58942 &MEDIUM \\         %V33 
RXJ0505.3--2849&05:05:20&-28:49:05&0.509$^{s}$&2001-09- 
01T13:17:36&0111160201  & 48867&THIN \\ 
RXJ0847.2+3449 &08:47:11&+34:49:16&0.560$^{v}$&2001-10- 
07T11:55:10&0107860501&81143&THIN \\            %V59 
RXJ1120.1+4318 &11:20:07&+43:18:05&0.600$^{b}$&2001-05-08T20:50:37&0107860201&  
22627&THIN \\ 
RXJ1325.5--3826 &13:25:20&-38:24:55 &0.445$^{s}$&2002-01-19T02:30:04  
&0110890101 &60894 &MEDIUM\\ 
RXJ1334.3+5030 &13:34:20&+50:30:54&0.620$^{b}$&2001-06-07T20:19:43&  
0111160101&47614&THIN \\ 
RXJ1354.2--0222 &13:54:17&-02:21:46&0.551$^{s}$&2002-07- 
19T15:16:53&0112250101&33374&THIN\\             %V151 
RXJ1701.3+6414 &17:01:23&+64:14:08&0.453$^{b}$&2002-05- 
31T17:49:42&0107860301&18172 &MEDIUM\\ \hline   %V190 
\end{tabular} 
\caption{\label{tab:observns} Summary of pointing directions for each observation, the overall 
scheduled durations, and redshift data ($^{s}$ Southern SHARC, \cite{southern};  
$^{b}$ Bright SHARC, \cite{brightsharc2}; $^{v}$ 160SD \cite{Vikh}).} 
\end{center} 
\end{table*} 
\normalsize 
 \subsection{XMM--Newton}
XMM-Newton (\cite{Jansen}) comprises 3 co-aligned telescopes, each
with effective area at 1.5keV of $\sim$1500cm$^{2}$, and Full Width
Half Maximum (FWHM) angular resolution of $\sim$5 arc-seconds. This
combination of the highest ever focused X-ray collection area, and the
ability to resolve clusters at all redshifts, makes XMM-Newton the
best suited observatory for this programme. The 3 telescopes each have
a focal plane CCD imaging spectrometer camera provided by the EPIC
consortium. Two also have a reflection grating array, which splits
off half the light, to provide simultaneous high resolution dispersive
spectra. These two telescopes are equipped with EPIC MOS cameras
(\cite{EPICMOS}), which are conventional CMOS CCD-based imagers 
enhanced for X-ray sensitivity. The third employs the EPIC PN
camera (\cite{EPICPN}) which is based on a pn-junction multi-linear
readout CCD.  The EPIC cameras offer a field of view (FOV) of $\sim$30
arc-minute diameter, and an energy resolution of typically 100~eV
(FWHM) in the range $\sim$0.2--10~keV. The 7 CCDs in each of the MOS
cameras are about 10 arc-minutes square each. The central chip
encompasses the whole of our $z>0.4$ clusters. The 12 CCDs in the PN
are about 4 $\times$ 13 arc-minutes.  Even if correctly centered on the
boresight PN CCD, some portion of the target cluster emission may spill
onto neighbouring CCDs, and across dead zones between CCDs.  In each
camera, an aluminized optical blocking filter was deployed, the
thickness of which was chosen to suit the expected brightness of
nearby serendipitous objects in each field. Seven of the eight
clusters were observed in the Full Frame Imaging mode, as appropriate
for weak extended targets, but one (RXJ1325.5) was observed when the
MOS cameras were in ``Window'' mode, due to the presence of an
unusually bright nearby point source.
 
The data were processed through the XMM-Newton Science Analysis 
Sub-System (SAS; \cite{SAS}) version 5.3, in order to register photons from 
detector to sky co-ordinates, to correct energy data for gain and 
charge transfer losses, and remove instrument noise artifacts. This 
provided calibrated event lists as a starting point for our detailed 
data reduction. 
 
\section{Data Reduction} 
 
\subsection{Rate Filtering}\label{sec:rate} 
 
\begin{figure} 
\begin{center} 
\resizebox{\hsize}{!}{\includegraphics{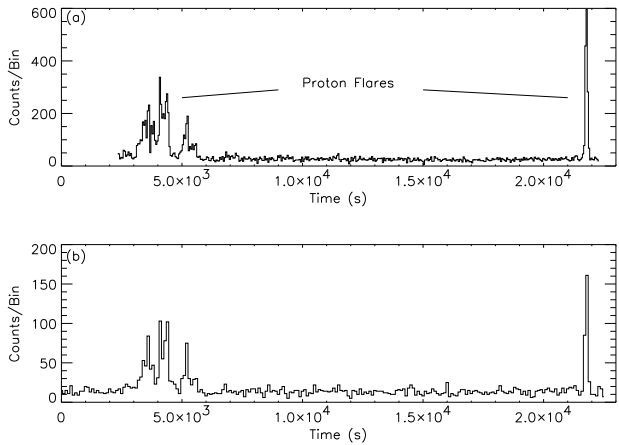}} 
\caption{Count rate in PN (top) and MOS (bottom) cameras in the RXJ1120.1 
  observation after selection for single pixel events above 
  10keV. Typical rates $\sim$15cts/50 sec bin (PN) and $\sim$8cts/100 
  sec bin (MOS) } 
\label{fig:bins} 
\end{center} 
\end{figure}

\begin{figure} 
\begin{center} 
\resizebox{\hsize}{!}{\includegraphics{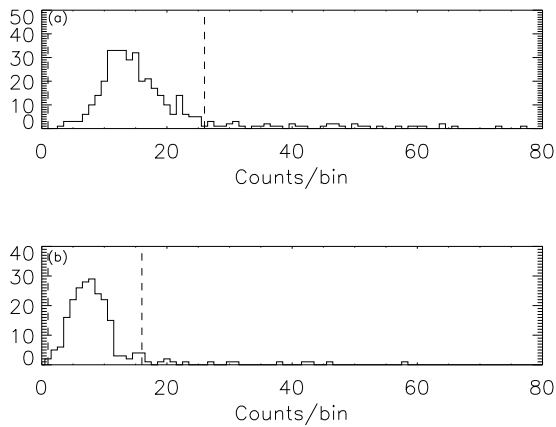}} 
 
\caption{Histogram of the count rate sequence for PN (top) and MOS
(bottom) cameras in the RXJ1120.1 observation after selection for
single pixel events above 10keV. The dashed lines indicate the upper
and lower ($\pm 3 \sigma$) bounds used for the {\em Good Time Interval}
filtering.}
 
\label{fig:binfit} 
\end{center} 
\end{figure}

Part of the XMM-Newton orbit lies within the magnetosphere, and
consequently the spacecraft can encounter clouds of protons
accelerated by magnetic reconnection. When these particles scatter
through the mirror system they are concentrated onto the focal plane,
and an enhanced background rate can occur. These intervals were
identified by forming histograms of events with energy $\geq$10~keV,
located in single pixels, in time bins of 50 (100) seconds in the PN
(MOS) camera(s). Due to variations in the baseline raw cosmic ray rate
experienced through the mission to date (probably due to modulation by
the solar activity and/or the seasonal variation of satellite apogee
direction), we prefer not filter at fixed background count rates, rather
for each exposure we define $\pm 3 \sigma$ limits after ignoring the
highest count rate periods (Figure~\ref{fig:bins} and
\ref{fig:binfit}).  Table~\ref{tab:expos} summarizes the total on-axis
{\it Good Time Interval} (GTI) exposure times extracted for each
observation using this procedure.
 
\begin{table}[t] 
\begin{center} 
\begin{tabular}{l c l l  l l}\hline\hline 
Cluster ID&MOS &PN &\% lost \\\hline 
RXJ0337.7--2522 & 23421 & 23611&60 \\ 
RXJ0505.3--2849&31851&24983&35\\ 
RXJ0847.2+3449 & 43793 & - &50\\ 
RXJ1120.1+4318 & 18333 & 16208&18\\ 
RXJ1325.5--3826 & - & 48753&20\\ 
RXJ1334.3+5030 & 40713 & 34909&15\\ 
RXJ1354.2--0222 & 9090& 7248&73\\ 
RXJ1701.3+6414 & 9379 & 4382&50\\\hline 
\end{tabular} 
\caption{\label{tab:expos} Summary of the usable exposure durations
  obtained for the different  clusters after accounting for the GTI
  filtering, telemetry losses etc.. In some observations data from only one camera was available. For operational reasons the PN camera commences exposures somewhat later than the MOS cameras. Losses due to proton flares, telemetry drops etc. are given in column 4 as a percentage of the total exposure time.} 
\end{center} 
\end{table} 

 For clusters  RXJ1354.2--0222 and RXJ1701.3+6414 the high prevalence of
  proton flares mitigated against the automated filtering procedure,
  and a manual analysis of the the ``curve of growth'' for
  signal:noise in the cluster target region was employed.
\subsection{Vignetting} 
\label{sec:vignet} 

The clusters in our sample are expected to be extended on a scale of
2--3 arc-minutes, and in order to assess accurately the surface
brightness distribution, a small correction for energy dependent
telescope vignetting must be applied. This was performed by the SAS
task {\it EVIGWEIGHT} which assigns a weighting value to each detected
photon, accounting for telescope and CCD efficiency variations (the
latter being negligible by comparison). Thereafter, spectrum
extraction and image product generation is automatically weighted for
the vignetting correction, and response matrices for the on-axis
location can be used. \footnote{In principle this may not apply
  correctly for objects {\em far} from the CCD array centre, as the detector response
  redistribution matrices may diverge slightly from the on-axis
  case. It is probably also true that mirror vignetting 
calibrations become less secure with increasing off-axis angle.}

An exposure map was also generated for each camera observation. These
maps accounted for spatial variations, such as dead pixels, noisy
readout columns and chip gaps, not included in the vignetting
correction. The XMM-Newton satellite is usually very stable during
observations, so the inter--CCD gaps remain approximately fixed in sky
projection.  It is therefore important to correct for flux lost from
 extended cluster regions  using the exposure maps. This is
especially true for the PN images, because the PN CCDs are smaller 
and the gaps closer to the target than those in the MOS cameras.
 
\subsection{Background subtraction techniques} 
\label{sec:back} 

Background subtraction is a very important step in our analysis.  Even
after the proton flare removal described above
(section~\ref{sec:rate}), and after masking out point sources, our
cluster observations will be contaminated, to varying degrees, by
three different background signals; ({\it i}) the cosmic X-ray
background, ({\it ii}) a particle background induced by incident
cosmic rays and ({\it iii}) some residual soft proton
contamination. We will refer to these as the cosmic, particle and
proton backgrounds respectively, hereafter.  There are two approaches
to background subtraction, either one can use the source observation
itself (this is known as {\it in-field} subtraction), or one can use
background template files. The templates are generated by combining
several deep, blank field, observations and are especially useful when
the source of interest covers a large fraction of the field of view
(e.g. a nearby cluster). We have used a combination of both techniques
in the analysis below.
 
For each of the eight clusters in our study, we applied the same rate
filtering, vignetting correction and detector-to-sky conversion used
for the reduction of the cluster observation, to the background
template files provided by the XMM-SOC (\cite{Back}). This allowed us
to extract identical {\em physical} detector regions for both the
cluster and the background using the SAS task $ATTCALC$.  We then
re-normalized the background template to account for any differences
in the particle background count rates between the cluster and
background template observations. This step was necessary because the
particle background varies with observation epoch
(Section~\ref{sec:protons}). To make this correction, we took
advantage of the fact that high energy particles are not focused by
the telescope optics, so the particle background count rate can be
measured from areas {\it outside} the telescope FOV.

We used re-normalized background templates (energy range 0.3--4.5~keV)
during the spatial analysis (Section~\ref{sec:radial}). We also used
background templates (energy range 0.3--10~keV) during the spectral
analysis of two of our two brightest clusters (RXJ1120.1 and
RXJ1334.3).  For this, we employed the so-called ``double
subtraction'' technique (\cite{doublesub}), which involves making an
additional correction to compensate for the fact that the Galactic
Halo and the Local Hot Bubble component of the diffuse cosmic X-ray
background varies significantly across the sky. To determine this
correction, we compared off-axis, source free, regions in both the
cluster fields and their corresponding background templates.  We found
the correction to be small, due to the lack of soft X-ray emission
features at the high Galactic latitude locations of RXJ1120.1 and
RXJ1334.3.  We have used the results from the double subtraction
analysis of RXJ1120.1 and RXJ1334.3 to validate the results from the
in-field background subtraction, see below.
 
For a variety of reasons, e.g. the choice of filter\footnote{The
\cite{Back} background templates are only appropriate for observations
made using the Thin filter; three of our clusters were observed
through the Medium filter (Table~\ref{tab:observns})}, low signal to
noise, non-zero off-axis angle etc., we were not able to apply the
double subtraction technique to all our clusters. Instead we have to
rely on the in-field technique.  The radial dependence of the
vignetting means that this technique is most successful for point
sources, however it should still work well for our clusters since, at
$z>0.4$, they are barely more extended than the instrument point
spread function (PSF). We illustrate this by considering a cluster
described by a $\beta$-model with   $\beta \sim$0.67 and a core
  radius of $25''$ ($\sim 185 h^{-1}_{50}$ kpc  typical values for the clusters in our sample,
Table~\ref{tab:L}). At a radius enclosing 75\% of the counts from such
a cluster, the vignetting is different by less than 1\% from the
aim-point value. Therefore we posit that an accurate representation of
the cosmic X-ray background spectrum at the cluster position can be
derived using a 
nearby in-field background aperture. For the spectral analysis of our
clusters we generally used an annular background aperture centered on
the source.  The annuli were chosen on a case by case basis, but were
typically $2-3'$ wide with an inner radius no less than $3'$ from the
cluster center (to ensure that no flux from outlying regions of the
cluster was erroneously removed). We note that in the case of
RXJ1325.5, we could not use an annulus, due to the proximity of a very
bright point source (Section~\ref{sec:1325}). In
Sections~\ref{sec:1120} and \ref{sec:1334}, we compare the results of
the spectral fitting for the clusters where both in field and double
subtraction could be applied. As we did not find significant
differences, we conclude that the in field background subtraction
technique should be valid for our sample.
 
\subsubsection{Impact of Residual Proton Contamination} 
\label{sec:protons} 

\begin{figure} 
\begin{center} 
\resizebox{\hsize}{!}{\includegraphics[bb= 50 10 500 340,clip]{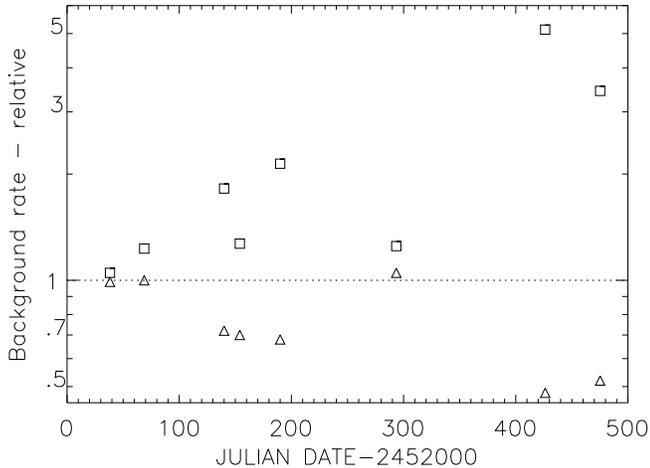}} 
 
\caption{\label{fig:cray} Comparison of the relative scaling of particle (triangle)  
and proton (square) background rates ({\em after GTI filtering}) in 
the eight cluster observations, with respect to the components of  the
background template files. Data averaged for all cameras in observation.} 

\end{center} 
\end{figure}

The ratio of the proton to particle backgrounds is not constant; even
after proton flare mitigation (section~\ref{sec:rate}), the proton
background may have a significantly different counting rate in the
cluster observation compared to background template files
(\cite{a1835}). This is illustrated in Figure~\ref{fig:cray}. The
triangles indicate the ratio of the particle background in the cluster
observation to that in the background template file (column 2 in
Table~\ref{tab:bkgd_ratio}). The squares show the equivalent
information for the proton background (column 3 in
Table~\ref{tab:bkgd_ratio}). For this comparison, the particle
background was estimated from the $\geq$10~keV count rate in CCD areas
{\it outside} the telescope FOV, whereas the proton background was
estimated by subtracting that value from the $\geq$10~keV count rate
within the FOV.  From the Figure~\ref{fig:cray}, it is clear that the proton
background varies with epoch and is anti-correlated with the particle
background. This anti-correlation can be explained if
\begin{itemize} 
\item  enhanced solar activity deposits more protons into the  
magnetosphere, expanding the latter and thus shielding the cosmic ray 
flux more efficiently.
\item  seasonal variations in orbit take the spacecraft in and out of the  
magnetosphere at apogee, so that cosmic ray flux shielding varies inversely  
with the exposure to magnetospheric protons.
\end{itemize} 
Although this anti-correlation tends to maintain a more constant {\em
  total} background rate, the result of this anti-correlation is that
  we cannot 
be certain the background
template re-normalizations described above, compensate correctly for
the temporal variations in the proton background. This suggests that
the in-field background subtraction method may be preferable to double
subtraction technique for {\em spectral} analysis.  However, the in-field
technique is complicated by the fact that the scattering of low energy
protons at the mirrors occurs over angles somewhat larger than for the
X-ray reflection (\cite{rasmussen}).  Therefore, when using task {\it
EVIGWEIGHT} (Section~\ref{sec:vignet}), there is the potential to
over-weight the proton background. The proton background spectrum is
hard, so this over-weighting could result in an artificial softening
of the cluster spectrum and hence a lowering the estimated cluster
temperature, which might in turn mimic evolution in the measured
$L_{\rm x}-T_{\rm x}$ relation (see section~\ref{sec:LT}).
 
\begin{figure} 
\begin{center} 
\resizebox{\hsize}{!}{\includegraphics{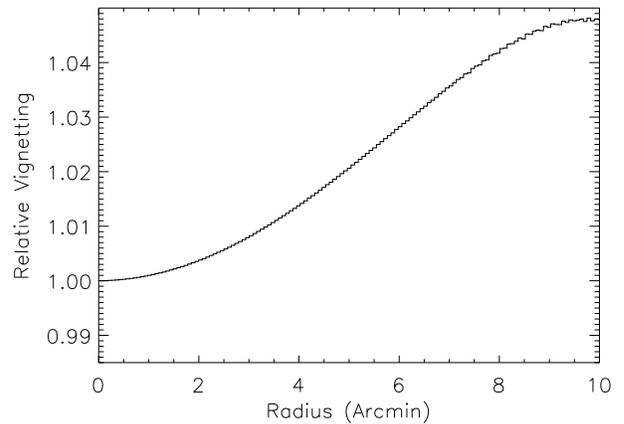}} 
\caption{Ratio of the surface brightness distributions in blank fields,  
showing  proton contamination in RXJ1701.3+6414 field.}
% and at  nominal fluxes } 
%??caption does not really make sense 
%%AB to check.
\label{fig:protvig} 
\end{center} 
\end{figure}

To examine the likely impact of systematic errors in the proton vignetting correction,
we used the {\it EVIGWEIGHT} vignetting corrected background template
files to create an image that was essentially free from proton
contamination. We did so by applying very strict count-rate
filtering. We then made a comparison image by applying less
conservative count-rate filtering to the background template
files. For this we used the filtering criteria derived from the
RXJ1701.3 observation (which suffered from unusually high proton
contamination); the comparison is presented in
Figure~\ref{fig:protvig}. 
The two count rate images we derived, should be
identical within the noise, except with regard to the proton
background. By dividing one by the other, and then fitting a radial
surface brightness profile, we should therefore be able to get an
impression of how the proton background is vignetted. The results of
this test (in the 0.5--7.0~keV band to emphasize an effect of hard
proton spectrum) are shown in
Figure~\ref{fig:protvig}. We also generated the equivalent figure
using the RXJ1354.2 filtering criteria with almost identical results.
From this investigation, we conclude that the proton background is
over-weighted by only a few percent by {\it EVIGWEIGHT} even at the
edges of the field of view in observations with significant proton contamination. We
therefore chose to ignore the over-weighting of the proton background
in our analysis, except in the cases of RXJ1701.3 and RXJ1354.2. For
these, we applied a small ($\sim$2\%) scaling to the background
spectrum derived from an in-field annulus around the cluster. 
\footnote{We caution the reader not to use Figure~\ref{fig:protvig} as a general template for XMM-Newton analysis, as both the spectrum and the absolute count rate of the proton background are likely time dependent.}

\begin{table} 
\begin{center} 
\begin{tabular}{l l l l l}\hline\hline 
Cluster ID&Outside FOV &Inside FOV\\\hline 
RXJ0337.7 ~--~2522 & 0.72&1.82   \\ 
RXJ0505.3 ~--~ 2849&0.70& 1.27\\ 
RXJ0847.2 + 3449 & 0.68&2.14\\ 
RXJ1120.1 + 4318 &0.99&1.05\\ 
RXJ1325.5 ~--~ 3826 & 1.05\\ 
RXJ1334.3 + 5030 &1.00&1.23\\ 
RXJ1354.2 ~--~ 0222&0.52&3.44\\ 
RXJ1701.3 + 6414 &0.48&5.13\\\hline 
\end{tabular} 
\caption{\label{tab:bkgd_ratio} The scaling ratio of the $>10$ keV count rates outside 
(column 2) and inside (column 3) the telescope FOV in the background 
template files compared to the cluster files. The former ratio is an 
indication of the particle (cosmic ray) background rate, while the 
latter indicates the level of the soft proton background rate. These ratios are 
plotted as a function of Julian date in Figure~\ref{fig:cray}} 

\end{center} 
\end{table} 
 
\subsection{Image Products} 
\begin{figure*} 
\begin{center} 
\resizebox{\hsize}{!}{\includegraphics{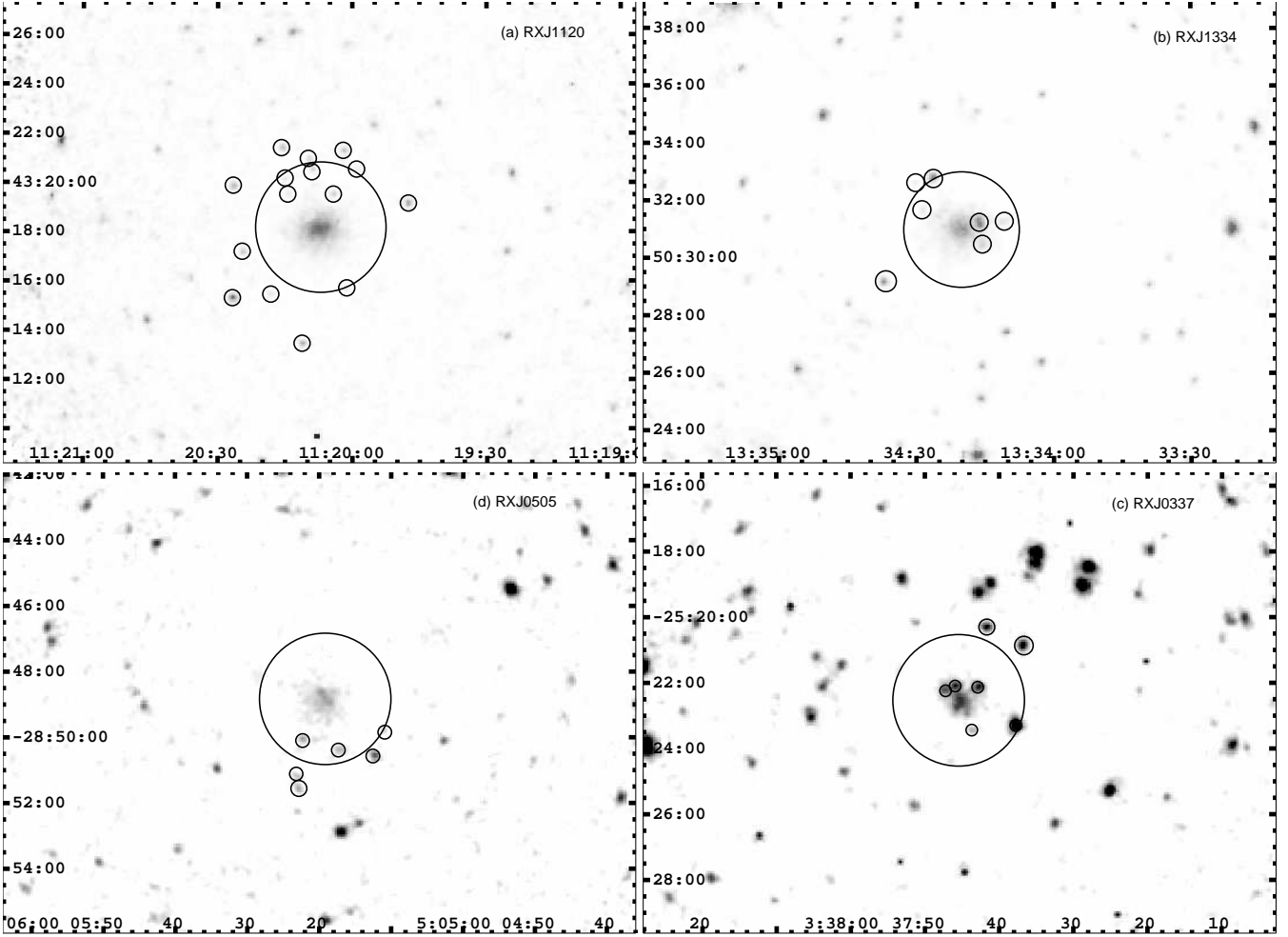}} 
\caption{\label{fig:image_i} 
(a) Central $\sim 18 \times18$ arc-minutes of the RXJ1120.1 field - smoothed
by a $4'' \sigma$ Gaussian. The large circle represents the spectral
extraction radius, and the smaller circles locate the point sources
which were excised. Similar images for RXJ1334.3 (b), RXJ0337.7 (c)
and RXJ0505.3 (d). The RXJ1334.3 image $15\times15$ arc-minutes; the
RXJ0337.7 and RXJ0505.3 images are $14\times14$ arc-minutes.}

\end{center} 
\end{figure*}

 Images for each camera were compiled in the 0.3--4.5~keV energy band,
this maximizes source count rates for cluster spectra that are
characterized by a $\sim$4keV temperature, and furthermore provides a
guard against corrections at the softest and hardest energy bands that
are subject to the largest potential Galactic and cosmic ray
subtraction inaccuracy. A spatial binning of 4.3 arc-seconds per pixel,
was employed, which slightly oversamples the mirror FWHM. The 3
separate count-rate images from the EPIC cameras were exposure
corrected and co-added. In each image field, we use the SAS task
$EBOXDETECT$ to identify point sources via a sliding box detection
algorithm (see Table~\ref{tab:srclist}). In Figures~\ref{fig:image_i}
and \ref{fig:image_ii}, we show the inner portion of the eight cluster
images after a Gaussian smoothing of $\sim 4''$ (using the $FTOOL$
$fgauss$). 
It is clear from these images that the clusters are well resolved and encompass a
variety of morphologies. Point sources masked during the spatial and
spectral analysis (section~\ref{sec:anal}) are indicated by small
circles. The larger circles denote the apertures used to generate the
cluster spectrum.
 
\subsection{Radial surface brightness profiles} 
\label{sec:radial} 
Before generating the profile, we defined the centroid of the cluster
brightness distribution using a 2-d Gaussian fit around the core of
the raw cluster image (0.3--4.5~keV). These centroids are
given in Table~\ref{tab:L}, row 1.  Next, a background correction was
applied by subtracting, pixel by pixel, the corresponding
re-normalized background template (Section~\ref{sec:back}). The radial
bins were chosen so that the background-subtracted counts per bin, in
the co-added profile, were at $\geq 3\sigma$ significance.  A
$\beta$--model convolved with the XMM-Newton telescope PSF appropriate
at the position of the cluster centroid was then fit to these
profiles, using simple $\chi^{2}$--minimization (background was
fixed). For the PSF
convolution, we used calibration file XRTn\_XPSF\_0004.CCF in SAS
medium accuracy mode, available from the XMM-SOC calibration ftp site.
The PSF was constructed by co-adding the contribution from PSFs at
different energies using a weighting scheme appropriate for a 4keV
thermal spectrum. Fortuitously the XMM-Newton telescope PSF is a
rather weak function of X-ray energy, so that any deviations from this
default spectrum would have a negligible impact on our fits.  We were
able to use the on-axis PSF for the convolution, except in the case of
RXJ1325.5. For this cluster, which was observed off-axis, the
appropriate off-axis PSF model was generated. We note that the PSF
correction was applied separately for each camera.  Fitted values for
$\beta$ and the core radius $r_c$ are given in Table~\ref{tab:L} rows
 8 \& 9. To convert the core radii from angles to distances for this
Table, we assumed a spatially flat cosmology with $\OmM=1$ and $H_0 =
50$~km~s$^{-1}$~Mpc$^{-1}$. The eight radial profiles and their
respective best fit PSF convolved $\beta$ models are presented in
Figures~\ref{fig:sb_i} and \ref{fig:sb_ii}.
 
\subsection{Spectral extraction} 
 
Cluster spectra  were extracted from circular regions centred on the
cluster centroids, with radii ranging from $90$ to $145$ arc-seconds
(typically $120''$, see Table~\ref{tab:L}, row 13). From these regions
we extracted data from the event list using the recommended pixel
selections. For the PN camera we used single and double pixel events
(event patterns 0--4) with all the recommended selection flags
applied. For the MOS cameras, we used the 0--12 event patterns. We
account for background contamination in the spectra using either the
double and/or in-field subtraction methods (Section~\ref{sec:back}).
 
Spectral fitting was implemented with the XSPEC v11 package 
(\cite{XSPEC}), using the MEKAL models (\cite{mekal}) for a thermal 
spectrum, modified with interstellar absorption (\cite{wabs}) 
appropriate for the Galactic column density (\cite{NH}).  To 
facilitate fitting via Chi-square minimization, the spectral files 
were re-binned to ensure at least 25 counts per bin in order to 
approximate Gaussian statistics. The energy range was 0.3-10keV.
 In general, the following on-axis 
response files were used during the fits, {\it 
epn\_ff20\_sdY9\_filter.rmf} for the PN camera and {\it 
m1(2)\_filterv9q20t5r6\_all\_15.rsp} for the MOS1(2) 
cameras. Exceptions to this, e.g. during the analysis of RXJ1325.5, 
are noted below. Fitted spectral parameters are quoted with 1$\sigma$ 
confidence limits on one interesting parameter. The spectral fits were 
typically performed simultaneously on the PN, MOS1(2) 
spectra. Again exceptions to this, e.g. during the analysis of RXJ0847.2, 
are noted in the relevant sections. 
 
Following Markevitch (1998), we have also investigated the impact of
cooling cores on our fitted temperatures, and hence the measured
$L_{\rm x}-T_{\rm x}$ relation, by performing spectral fits after
excising the central region. For all clusters except RXJ1701 and RXJ1325, a 50
h$^{-1}_{50}$ kpc region was excised. For RXJ1701 \& 1325 a 120kpc region was excised (see
section 4.7). The results from these
fits are given at the beginning of Table~\ref{tab:L}. See Section~\ref{sec:cflows}
for a discussion.
 
\subsection{Deriving Bolometric Luminosities} 
\label{sec:Lx-measure} 

%??are the fluxes in Table 5 absorbed or unabsorbed?
For the luminosity calculations, we adopted a physically meaningful
circular aperture, with a virial radius, $r_v$. We derived $r_v$
according the $T-r_v$ relation of Evrard et al. (1996) for each cluster using
the best fit temperatures from the spectral fits. We note that
uncertainties both in the measured temperature and in the $T-r_v$
relation will introduce a systematic error in the derived $L_{\rm x}$
values. However, this should be insignificant, as very little cluster
flux falls at radii close to the virial radius.  

We integrated the
background-subtracted counts inside $r_v$ and then corrected for any
cluster flux lost in areas under masked point sources or inter-chip
gaps. The correction factors are given in  row 12 of
Table~\ref{tab:L}. These were derived in a model independent fashion;
the count-rate was accumulated in 1 pixel wide annuli centered on the
cluster. If any pixels within a specific annulus fell in chip gaps or
in the point source masked, the annulus count-rate was scaled up
accordingly. Next, we used XSPEC, together with the appropriate EPIC
response functions, to determine the absorbed 0.3-4.5 keV flux within
$r_v$, using the best fit absorbed MEKAL model from our spectral
analysis so that it yielded the measured $r<r_v$ count rate (the MEKAL
normalization factors are given in row 10 of Table~\ref{tab:L}). 
For comparison with equivalent ROSAT data such as SHARC or 160SD these
fluxes (converted to the normal ROSAT 0.5--2~keV band) are listed in
row 6 of Table~\ref{tab:L}.  We
then set the hydrogen column density to zero and re-determined the
flux inside a pseudo bolometric band of 0.0--20 keV. From the
unabsorbed bolometric flux, it is trivial to determine the bolometric
luminosity for any given cosmological model. To aid comparisons with
previous work, we give in, row 3 of Table~\ref{tab:L}, the $L_{\rm x}$
values appropriate for a  $\OmM=1$, $\OmL=0$,
$q_0=0.5$ model with $H_0=50$~km~s$^{-1}$~Mpc$^{-1}$ that has been
most frequently used in the past as the parameter set for X-ray cluster studies.
 
 \begin{figure*} 
\begin{center} 
\resizebox{\hsize}{!}{\includegraphics{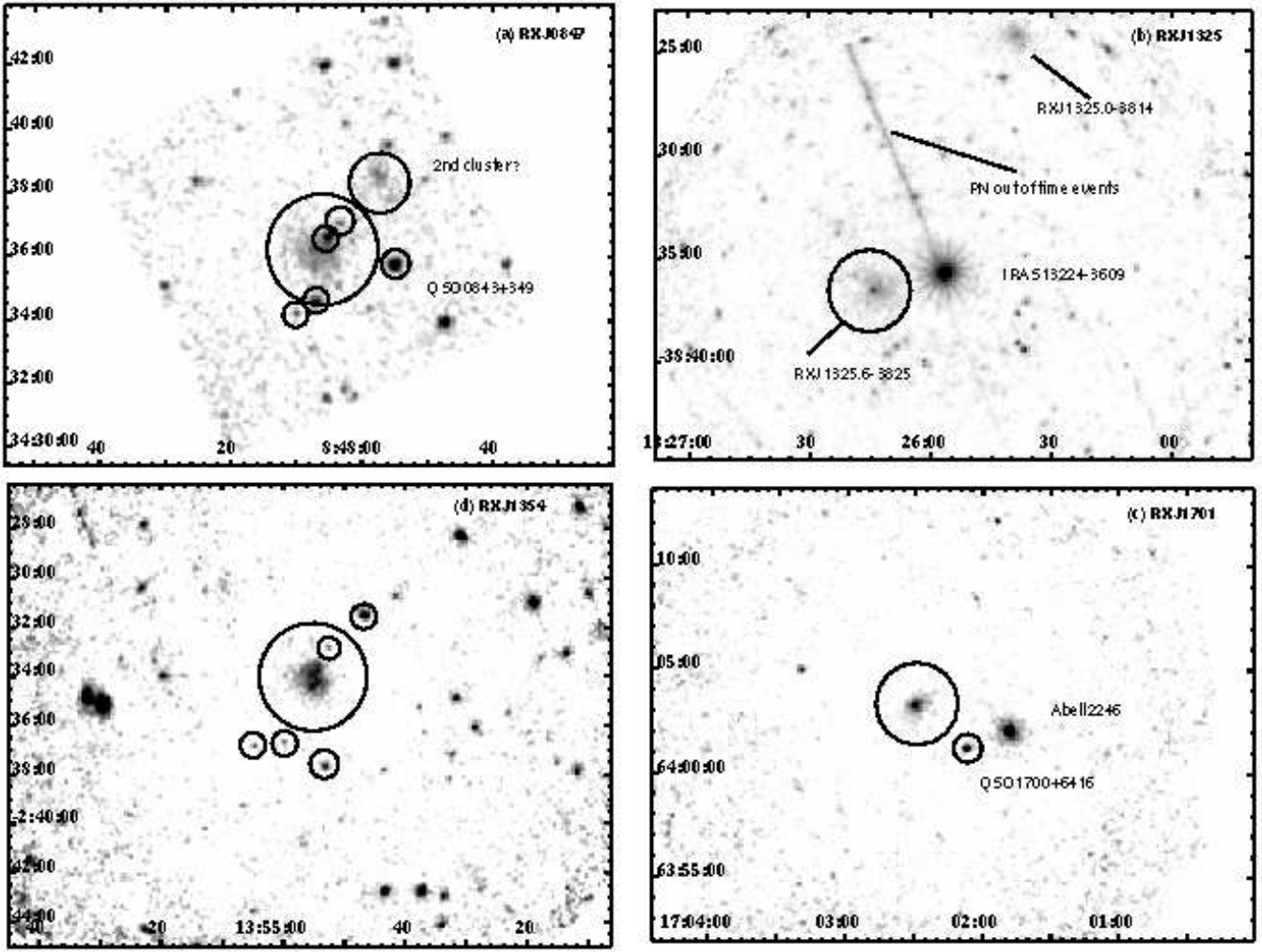}} 
%??modified to reflect order in text; sigma or FWHM on Gaussian?
%??Do I have the image sizes correct?

\caption{\label{fig:image_ii}  
(a) Central MOS CCD image (10 arc-min) of the RXJ0847.2 field - smoothed by a $4''$
Gaussian. The large circle represents the spectral extraction radius,
and the smaller circles locate the point sources which were
excised. Similar images for RXJ1325.5 (b), RXJ1701.3 (c) and RXJ1354.2
(d). A newly discovered candidate distant cluster (XMMU
J084701.8+345117) is labeled to the upper right of RXJ0847.2.  The
RXJ1325.5 and RXJ1325.0 Southern SHARC clusters fall in an observation
centered on IRAS 13224-3809. The long steak connected to IRAS
13224-3809 comes from the "out-of-time" events collected during
readout of the PN camera.  In the RXJ1701.3 field, the nearby Abell
2246 and QSO objects are labeled. The RXJ1325.5  image is  $\sim
18\times18$ arcminutes; the RXJ1701.3 and RXJ1354.2 images are $\sim 14\times14$
arcminutes.}
\end{center} 
\end{figure*} 

Also following Markevitch (1998), we estimated the cooling flow
corrected bolometric luminosities by summing counts in a (r$_{min}
\leq r \leq r_{v}$) ring. For clusters other than RXJ1701 and RXJ1325, r$_{min}$ =
50 h$^{-1}_{50}$ kpc, while for RXJ1701\&1325  r$_{min}$=120 h$^{-1}_{50}$
kpc (see section 4.7). To account for the non-cooling flow flux
falling at r$\leq r_{min}$, we applied a cluster specific
renormalisation parameter calculated using the best fit
$\beta$-model. We note that our correction approach differs from that
of Markevitch (1998), therein a single multiplicative factor of 1.06
was used.
\section{Individual Clusters}\label{sec:anal} 
 
We now discuss our analysis for each cluster in turn. We have ordered
our discussion according to the Julian date on which they were
observed (Table~\ref{tab:observns}).
 
%?? general comment: mention the angular limits of the beta-model  
%fit for each cluster 
 
\subsection{RXJ1120.1+4318} \label{sec:1120} 
 
Figure~\ref{fig:image_i}(a) shows the vignetting corrected, background
subtracted and co-added (PN+MOS1+MOS2) 0.3--4.5~keV image of
RXJ1120.1+4318. Point sources removed during analysis are circled and
their positions listed in Table~\ref{tab:srclist}.
Figure~\ref{fig:sb_i} shows the corresponding radial surface
brightness distribution and best fit $\beta$ model;
$\beta=0.77~\pm~0.03$, $\theta_c=27.4~\pm~1.2$ arcseconds
($r_c=209~^{+9}_{-8}~ h^{-1}_{50}$ kpc, Table~\ref{tab:L}).
 
From a spectral extraction region with a radius of $145''$ and using 
the in-field background subtraction technique, we measured the 
following temperature, metal abundance and redshift by fitting to the 
PN, MOS1 and MOS2 data simultaneously; $T_{\rm x}= 5.45 ^{+0.26}_{-0.35}$ 
keV, $Z=0.47 \pm 0.09$, $z=0.60 \pm 0.08$. We note the consistency between the redshift 
measured from the X-ray spectrum with the optically determined value 
($z=0.60$; Romer et al. 2000). The overall $\chi_{\nu}$ of the 
spectral fit was 390 / 360 degrees of freedom. During the fit, the 
hydrogen column density was fixed at the Galactic value 
(\nh=$2.1\times 10^{20}$ atoms cm$^{-2}$), but we note that, when 
left as a free parameter, its best fit value was very similar 
(\nh$=2.2 \pm 0.4 \times 10^{20}$ atoms cm$^{-2}$). 
 
\begin{figure} 
\begin{center} 
\resizebox{\hsize}{!}{\includegraphics[angle=270,scale=0.4]{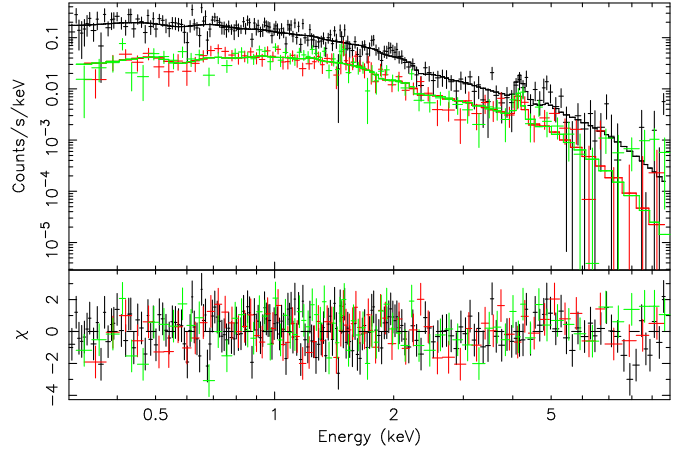}} 
\caption{\label{fig:rxj1120spec} Spectral fit and residuals for
RXJ1120.1; black - PN, dark and light grey - MOS1\&2 (double background
subtraction).}

\end{center} 
\end{figure}

Of the eight clusters in our sample, this object has the greatest
signal-to-noise ratio. We have taken advantage of this to perform
additional spectral analyses. First we have been able to use the
double subtraction technique to investigate how the adopted background
subtraction method impacts the spectral fits. Doing so, we derive
$T_{\rm x}=5.6^{+0.25}_{-0.3}$~keV, $Z=0.43 \pm 0.06$, $z=0.605 \pm
0.08$ (Figure~\ref{fig:rxj1120spec}).  The overall $\chi_{\nu}$ of the
spectral fit was 330 / 307 degrees of freedom. During the fit, the
hydrogen column density was again fixed at the Galactic value
(\nh$=2.1\times 10^{20}$ atoms cm$^{-2}$), but when left as a free
parameter, its best fit value was (\nh$=2.2 ^{+0.2}_{-0.4}\times
10^{20}$ atoms cm$^{-2}$).  It is clear, therefore, that the spectral
fits are not significantly changed by the choice of background subtraction
technique (see also section~\ref{sec:1334}).
 
\begin{figure} 
\begin{center} 
\resizebox{\hsize}{!}{\includegraphics[scale=.8,bb=60 15 490  
330,clip]{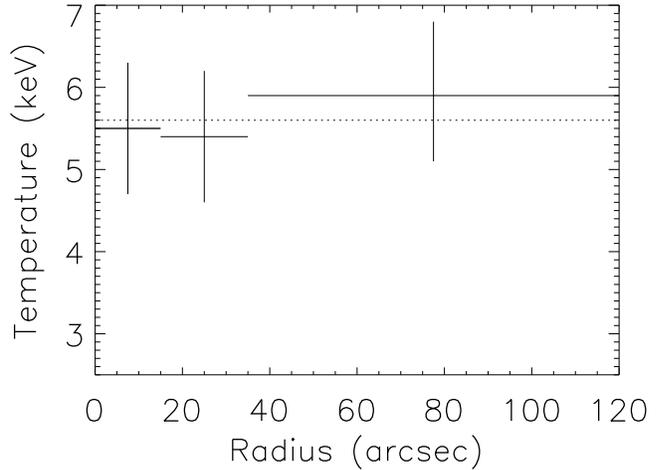}} 
\caption{The spatially resolved temperature profile of RXJ1120}
\label{fig:rxj1120tempprof} 
\end{center} 
\end{figure}

We have also investigated how the spectral fits differ when we treat
the PN and MOS data separately. We find them to be in good
agreement; $T_{\rm {x},PN}=5.30 \pm$ 0.6 keV and $T_{\rm {x},MOS}=5.7
\pm$ 0.8 keV. This is encouraging since, in two other cases (RXJ1325.5
and RXJ0847.2), we do not have access to data from all three cameras.
We have also been able to subdivide the spectral extraction region
into three radial bins, and determined a crude temperature profile
(Figure ~\ref{fig:rxj1120tempprof}). The profile is essentially flat,
indicating that there is not a ``cooling flow'' region at the cluster
core (this conclusion is supported by the absence of a central spike
in the surface brightness profile). We note that for the PN/MOS
comparison and for the radial profile, we used the double subtraction
technique to account for the background.
 
\subsubsection{Comparison with Arnaud et al. 2002} 
\label{sec:arnaud02} 

The RXJ1120.1 observation described here have been previously analyzed
and interpreted in an earlier paper in this series
(\cite{Arnaud}). The analysis procedures developed in that paper have
formed the basis of the analysis of the eight clusters described
here. For example, the vignetting correction technique used in
Arnaud et al. (2002) has been subsequently implemented in SAS as the {\it
EVIGWEIGHT} routine mentioned above. However, there are certain
differences in our respective data analysis techniques, even with
regard to the RXJ1120.1 observation. For example, we use updated
calibration information and an updated SAS processing version. We also
used a different method (3-sigma clipping) for the GTI filtering. We
have also adopted the in-field background subtraction method as the
standard for our cluster analysis (in Arnaud et al. (2002), only the double
subtraction technique was used). Finally, we note our use of the both
single and double PN events, compared to the selection of single
events by Arnaud et al. (2002). In general the revised analysis of
RXJ1120.1 has yielded very similar results to Arnaud et al. (2002) and we
concur that, given its isothermal temperature profile and the absence
of significant substructure, RXJ1120.1 is consistent with being a
relaxed cluster. It is noteworthy, however, that in Arnaud et al. (2002) the
mean temperature values determined using only MOS data differed from
those determined using only PN data by more than 1 keV; $T_{{\rm
x},MOS}=5.8^{+1.0}_{-0.7}$ keV and $T_{{\rm x},PN}=4.5^{+0.8}_{-0.5}$
keV. By contrast, we measured temperatures that differed by only 8\%
(see above). This improvement is most likely attributable to the
improved calibration that has become available since Arnaud et al. (2002) was
published. For completeness we compare in Table~\ref{tab:arn-lumb} the
values derived for various fitted parameters in the two analyses. For
consistency with Arnaud et al. (2002), we quote the mean temperature derived
using the double subtraction technique (in Table~\ref{tab:L} we quote
the value from  based on the in field subtraction technique).
 \begin{table} 
\begin{center} 
\begin{tabular}{ll ll}\hline\hline 
 $T_{\rm mean}$ & EdS $L_{\rm x}$  & $\beta$ & $\theta_c$ ($''$)\\  
                & ($10^{44}$ erg s$^{-1}$) & & \\ \hline 
 $5.3\pm0.5$ keV          & $13.9\pm{0.8}$ 
                                               & $0.78^{+0.06}_{-0.04}$ 
                                                         &$26.4^{+3.6}_{-2.4}$\\ \hline 
 $5.6^{+0.25}_{-0.3}$ keV & $14.4\pm{0.2}$  
                                               & $0.77\pm{0.03}$  
                                                         &$27.4\pm1.2$\\ \hline 
\end{tabular}

\caption{\label{tab:arn-lumb} Comparison of parameter fits to the
RXJ1120.1 observation by Arnaud et al. (2002) (first line) and this work
(second line). Here $L_{\rm x}$ refers to the bolometric luminosity.}
\end{center} 
\end{table}  
 
 
\subsection{RXJ1334.3+5030} \label{sec:1334} 
 
Figure~\ref{fig:image_i}(b) shows the image of the
RXJ1334.3+5030 field. Point sources removed during analysis are circled and
their positions listed in Table~\ref{tab:srclist}. 
 Figure~\ref{fig:sb_i} shows the corresponding
radial surface brightness distribution and best fit $\beta$ model;
$\beta=0.66~\pm~0.02$, $\theta_c=20 \pm1 $ arcseconds
($r_c=154\pm10 h^{-1}_{50}$ kpc, Table~\ref{tab:L}).

\begin{figure} 
\begin{center} 
\resizebox{\hsize}{!}{\includegraphics[angle=270,scale=0.5]{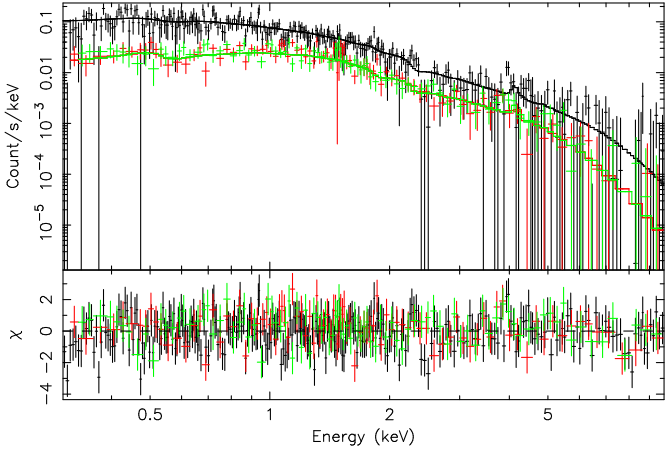}} 
\caption{\label{fig:rxj1334spec} Spectral fit and residuals for
RXJ1334.3; black - PN, dark \& light gray - MOS1\&2 (in-field background subtraction).}
\end{center} 
\end{figure}

From a spectral extraction region with a radius of $120''$ and using
the in-field background subtraction technique, we measured the
following temperature and metal abundance; $T_{\rm x}=5.20^{+0.26}_{-0.28}$~keV,
$Z=0.15 \pm 0.08$ (Figure~\ref{fig:rxj1334spec}).  The overall
$\chi_{\nu}$ of the spectral fit was 464 / 473 degrees of freedom.
The hydrogen column was fixed at the Galactic
value (\nh$=1.05\times 10^{20}$ atoms cm$^{-2}$), but we note that,
when left as a free parameter, its best fit value was very similar
(\nh$=0.8\pm 0.5 \times 10^{20}$ atoms cm$^{-2}$). Likewise, the
redshift was fixed at its optically derived value of $z=0.62$ (Romer
et al. 2000), but when left as a free parameter, its best fit value
was $0.63 \pm0.02$. The derived values for the bolometric luminosity
and absorbed flux inside the viral radius are given in
Table~\ref{tab:L}. We were also able to apply the double subtraction
technique to these data and derived a consistent mean temperature
value; $T_{\rm x}=5.05\pm0.3$~keV.
 
\subsection{RXJ0337.7-2522} 
 Figure~\ref{fig:image_i}(c) shows the vignetting corrected, background
subtracted image of the RXJ0337.7-2522 field.  Two excised sources within
the spectral extraction radius, XMMU J033742.9-252208 \& XMMU
J033745.9-252206, are tentatively associated with two stellar objects
on the USNO Catalogue.  The former is U0600-01432100, also identified
as blue stellar object PHL4470, the latter is identified as
U0600-01432383 (a 16.5 $r$ magnitude object).  Figure~\ref{fig:sb_i}
shows the radial surface brightness distribution and
best fit $\beta$ model; $\beta$=0.76$^{+0.08}_{-0.04}$ and
r$_{c}$=19.4~$\pm2.5$ arcseconds.  ($r_c=145\pm18 h^{-1}_{50}$ kpc,
Table~\ref{tab:L}).
  
\begin{figure} 
\begin{center} 
\resizebox{\hsize}{!}{\includegraphics[angle=270,scale=0.5]{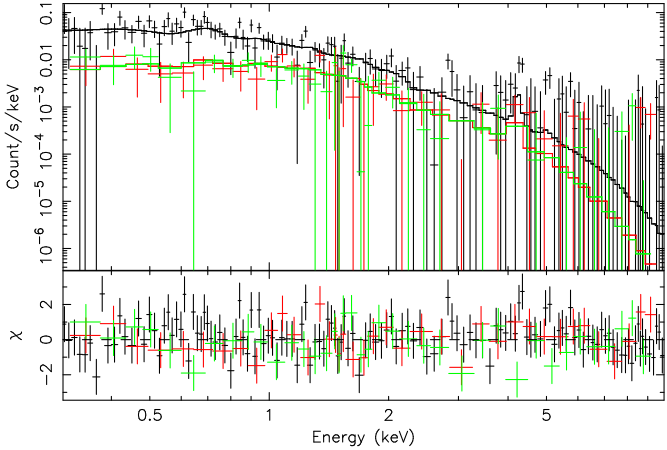}} 
\caption{\label{fig:sharc4spec} Spectral fit and residuals for
 RXJ0337.7; black - PN, dark\&light grey - MOS1\&2}
\end{center} 
\end{figure} 
 
From a spectral extraction region with a radius of $120''$ and using 
the in-field background subtraction technique, we measured the 
following temperature, metal abundance and redshift; 
$T_{\rm x}=2.6 \pm~0.35$~keV, 
$Z=0.38\pm 0.09$, $z=0.57\pm0.3$ (Figure~\ref{fig:sharc4spec}).  The 
overall $\chi_{\nu}$ of the spectral fit was 193 / 214 degrees of 
freedom.  We note the consistency between the redshift measured from 
the X-ray spectrum with the optically determined value ($z=0.577$; 
\cite{southern}). The hydrogen column density was 
fixed at the Galactic value (\nh$=0.99\times 10^{20}$ atoms 
cm$^{-2}$),  when left as a free parameter, its best 
fit value was very similar (\nh$=8.7^{+6.4}_{-5.4} \times 10^{19}$ 
atoms cm$^{-2}$).  The derived values for the bolometric luminosity 
and absorbed flux inside the viral radius are given in 
Table~\ref{tab:L}.

\subsection{RXJ0505.3-2849} 
 
The RXJ0505.3-2849 field is shown in Figure~\ref{fig:image_i}(d). 
 Four point sources removed during analysis are circled
and their positions listed in Table~\ref{tab:srclist}.  Tentative
identifications of these sources with objects in the UK APM survey
(\cite{UKAPM}) suggest an absolute astrometric accuracy for our
observations of $\sim$2 arcseconds. 
Figure~\ref{fig:sb_i} shows the corresponding radial surface
brightness distribution and best fit $\beta$ model;
$\beta$=0.66$^{+0.05}_{-0.04}$ and r$_{c}$=22.8~$\pm2.4$ arcseconds.
($r_c=164\pm17 h^{-1}_{50}$ kpc, Table~\ref{tab:L}).
  
\begin{figure} 
\begin{center} 
\resizebox{\hsize}{!}{\includegraphics[angle=270,scale=0.5]{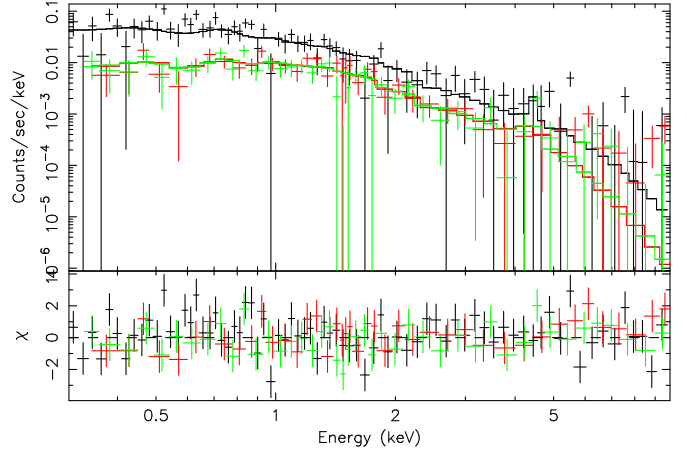}} 
\caption{\label{fig:sharc2spec} Spectral fit and residuals for
 RXJ0505.3; black - PN, dark \& light grey - MOS} 
\end{center} 
\end{figure}

From a spectral extraction region with a radius of $120''$ and using
the in-field background subtraction technique, we measured; 
$T_{\rm x}=2.5~\pm0.3$~keV, $Z=0.17 \pm
0.08$ (Figure~\ref{fig:sharc2spec}).  The overall $\chi_{\nu}$ of the
spectral fit was 279 / 248 degrees of freedom.  The fixed Galactic hydrogen
value was \nh$=1.5\times 10^{20}$ atoms cm$^{-2}$), compared with a
free 
fit parameter, of \nh$=1.1\pm 0.6 \times 10^{20}$ atoms
cm$^{-2}$). The 
redshift was fixed at its optically derived value of $z=0.51$ (\cite{southern}), but when
left as a free parameter, its best fit value was $z=0.53 \pm0.04$. 
As a test, we have
also performed a spectral fit without excluding the four point
sources.  We find that, within the errors, the fitted temperature was
unchanged.
  
 \subsection{RXJ0847.2+3449 (and XMMU J084701.8+345117)} 
Figure~\ref{fig:image_ii}(a) shows the image
of RXJ0847.2+3449. Of the eight clusters in our sample,
RXJ0847.2+3449 is noteworthy as being the only one that is not a SHARC
cluster; it was selected instead from the 160SD (\cite{Vikh}).
RXJ0847.2 was observed during a period when the PN camera was
temporarily disabled by a hardware fault, and the requested exposure
duration was partly compensated by an extension of the MOS
observation.  
To improve the contrast of the cluster against the nearby
bright QSO (PG 0844+34), only the inner part of the image, that
covering central CCD of the MOS cameras, is shown in
Figure~\ref{fig:image_ii}(a). We note that the mirror scattering is
sufficiently low that no flux from PG 0844+34 impedes our
analysis. The bright point source $\sim$3 arcminutes west of the
RXJ0847.2 cluster in Figure~\ref{fig:image_ii}(a) was identified with
a second, fainter, quasar, QSO 0843+349 at $z=1.57$.  The image also
shows evidence for a nearby {\em fainter} extended object to the NW,
which we identify as a cluster candidate with provisional designation
XMMU J084701.8+345117. Vikhlinin (private communication) claims to
have found a concentration of faint galaxies in this region,
but to date no galaxy redshifts are available. Neither QSO 0843+349
nor XMMU J084701.8+345117 fall inside our spectral extraction region.
The source detection software picked out four sources in that region,
none of which have counterparts in the NED catalogs. These sources
were excluded from the spatial and spectral analysis, they are circled
in Fig.~\ref{fig:image_ii} and their positions are listed in
Table~\ref{tab:srclist}.  

\begin{figure} 
\begin{center} 
\resizebox{\hsize}{!}{\includegraphics[angle=270,scale=0.5]{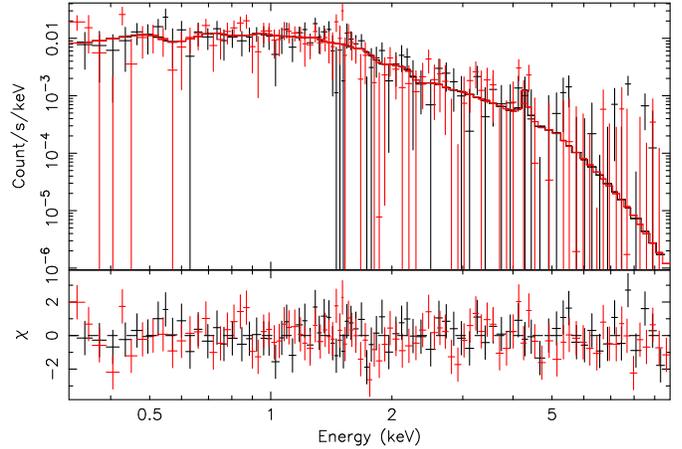}} 
\caption{\label{fig:vik59spec} Spectral fit and residuals for
 RXJ0847.2 (MOS data only)}
\end{center} 
\end{figure}

Figure~\ref{fig:sb_ii} shows the radial surface brightness
distribution for RXJ0847.2 and the best fit $\beta$ model;
$\beta$=0.81$\pm 0.07$ and r$_{c}$=42~$^{+4}_{-4}$ arcseconds.
($r_c=307\pm30 h^{-1}_{50}$ kpc, Table~\ref{tab:L}).
%Update the figure 
From a spectral extraction region with a radius of $120''$ and using
the in-field background subtraction technique, we measured 
$T_{\rm x}=3.62^{+0.58}_{-0.51}$~keV,
$Z=0.30 \pm 0.28$ (Figure~\ref{fig:vik59spec}).  The overall
$\chi_{\nu}$ of the spectral fit was 152 / 180 degrees of freedom.
The fixed hydrogen column density was \nh$=3.2\times 10^{20}$ atoms
cm$^{-2}$, 
as a free parameter, its best fit value was \nh$=2.8\pm 0.14 \times
10^{20}$ atoms cm$^{-2}$. The redshift was fixed at its optically derived value of $z=0.56$
but its best fit value
was $z=0.54 \pm0.04$. As a test, we have also performed a spectral fit
without excluding the four point sources. Doing so changed the fitted
temperature by 0.1~keV and the measured flux by $\leq$10\%.

We have also estimated the temperature of the second cluster in the
field, XMMU J084701.8+345117. For this we used a spectral extraction
radius of 80 arcsec and the in field background subtraction technique. We do
not have an independent estimate of the cluster redshift, and it
cannot be constrained by the X-ray spectrum, so we have assumed that
the cluster lies at the same redshift as RXJ0847.2 ($z=0.56$). We have
also fixed the metal abundance to be $Z=0.3$  and the hydrogen
column density to the same Galactic value. We measure a cluster
temperature of $T_{\rm x}=1.8^{+1.0}_{-0.4}$ keV and a corresponding
($r<r_v$, 0.5--2~keV) flux and bolometric luminosity of $1.3\times~10^{- 14}$ 
\ergcms and $1\times~10^{43}$\ergs respectively. These results are
consistent with measured low red-shift $L_{\rm x}-T_{\rm x}$ relations.

\subsection{RXJ1325.5--3826 (and RXJ1325.0-3814)} 
\label{sec:1325} 

The data for this cluster were obtained by kind permission of
Guaranteed Time holder M Watson, from an observation of the nearby
object IRAS 13224-3809.  Unfortunately for our own purposes, this
bright source had been observed in the MOS cameras in a ``Window''
mode of readout on the central CCDs in order to minimize effects of
photon pile-up, and consequently the area of focal plane containing
the SHARC cluster was not exposed. Therefore, our spatial and spectral
analysis is restricted to the PN camera data
only. Figure~\ref{fig:image_ii}(b) shows the vignetting corrected,
background subtracted 0.3--4.5~keV PN image of the region surrounding
IRAS 13224-3809. Clusters RXJ1325.5--3826 and RXJ1325.0-3814 (see
below) are labeled, as is the streak corresponding to the ``out of
time'' events from IRAS 13224-3809.  IRAS 13224-3809 is bright enough
that the ``spider'' like structure in the PSF is visible around
it. Given the proximity of the cluster to the IRAS source, the surface
brightness profile for RXJ1325.5--3826 (Figure~\ref{fig:sb_ii}) was
extracted from manually selected clean regions that avoided the IRAS
13224-3809 PSF. 

For the $\beta$ model fitting, a PSF appropriate for
the PN at the off-axis angle of the cluster was used (for all the
other clusters in our sample, we had used the on-axis PSF). The best fit
parameters in the spatial fit are $\beta=0.64^{+0.09}_{-0.07}$ and
$\theta_c=17.3 \pm 3.0$ arcseconds ($r_c=115 \pm 20 h^{-1}_{50}$ kpc,
Table~\ref{tab:L}).  There is evidence for possible excess brightness
in the inner bins of the radial profile. This may be in an indication
of a central cooling core, but we do not have adequate signal to noise
to confirm this spectroscopically. Following the procedure adopted for
RXJ1701.3 (see below) we also performed a $\beta$ fit after exclusion
of the core 120kpc and found a value for $\beta$ of 0.71$\pm$0.05,
given a fixed r$_{c}$ of 200$h_{50}^{-1}$kpc (We fixed this value
arbitrarily to match typical cluster values).

\begin{figure} 
\begin{center} 
\resizebox{\hsize}{!}{\includegraphics[angle=270, scale=0.5]{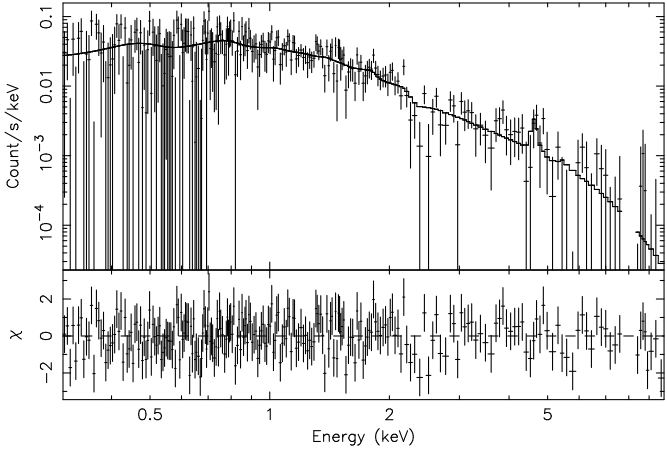}} 
\caption{\label{fig:rxj1325spec}  Spectral fit and residuals for
 RXJ1325.5 (PN data only)}
\end{center} 
\end{figure}

We used a $90''$ radius extraction region to fit the spectrum from
this cluster. This is a smaller region than was used for the other
clusters ($120''$ or $145''$, Table~\ref{tab:L}) because of the
proximity of IRAS 13224-3809.  Even though the proton background was
low enough in this exposure to permit spectral measurements using the
double subtraction technique, we decided to use instead only the
in-field background subtraction technique. This was because we wanted
to mimic the off-axis angle dependence of the point source
contamination.  Rather than using an annulus around the cluster to
determine the background, we used two source and streak free circular
regions at the same off-axis angle as the cluster.  From the
background subtracted PN spectrum, we measured the following
temperature, metal abundance and redshift; $T_{\rm
x}=4.15^{+0.4}_{-0.3}$~keV, $Z=0.31 ^{+0.19}_{-0.17}$, $z=0.44\pm{0.01}$
(Figure~\ref{fig:rxj1325spec}). We note the consistency between the
redshift measured from the X-ray spectrum with the optically
determined value ($z=0.445$; \cite{southern}). The overall
$\chi_{\nu}$ of the spectral fit was 229 / 242 degrees of
freedom. The fixed and free hydrogen column density values were
d(\nh$=4.8\times 10^{20}$ atoms cm$^{-2}$ and 4.5$^{+0.9}_{-0.8}
\times 10^{20}$ atoms cm$^{-2}$ respectively.

\subsubsection{RXJ1325.0-3814} 
 For completeness, we note that another, fainter, Southern SHARC
cluster lies within the same field of view. The PN image of this
cluster, RXJ1325.5-3814, can be seen in
Figure~\ref{fig:image_ii}(b). The cluster is also visible in
the MOS data (not shown) because it lies outside the region blanked by
the ``Window'' mode.  The cluster is too faint and too far off axis
({\it i.e. where the PSF is poorly defined}) to permit a detailed
spatial analysis, however we were able to make a spectral analysis. For
this we generated off-axis response matrices for each camera using the
SAS {\it ARFGEN} package.  From a spectral extraction region with a
radius of $90''$, and using the in-field background subtraction
technique, we measured the following temperature and redshift (\nh fixed at the
same galactic value of  4.8$\times 10^{20}$ atoms cm$^{-2}$ and abundance  at 0.3) $T_{\rm x}=3.2
\pm0.4$ keV and a redshift of $z=0.29 \pm0.02$.  We note the
consistency between the redshift measured from the X-ray spectrum with
the optically determined value ($z=0.296$; \cite{southern}). 

The cluster is observed close to the edge of the outer CCDs,
preventing the use of an extraction region large enough to correspond
with the virial radius. Within a
radius $\sim$500 $h_{50}^{-1}$kpc, the measured flux and bolometric luminosity are 6.1
($\pm$0.3)~10$^{-14}$ \ergcms (0.5--2~keV) and $L_{\rm x}=1.1 \pm 0.23
~10^{44}$\ergs respectively.  This combination of $T_{\rm x}$ and
$L_{\rm x}$ are not consistent with the measured $L_{\rm x}-T_{\rm
x}$ relation for nearby clusters. For example, for $T_{\rm x}=3.2$
keV, one would expect a bolometric luminosity of $L_{\rm x}= 2 \times
10^{44}$\ergs based on the Markevitch (1998) relation, and is probably
due mainly to the missing flux outside our spectral extraction radius.
  
\subsection{RXJ1701.3+6414 (and Abell 2246)} 
 
This cluster was observed during an exposure scheduled near the end of
an orbit, where the spacecraft was approaching the particle belts. Not
only were the spacecraft operations terminated prematurely, but most
of the exposure was dominated by high soft proton background from the
edge of the particle belts. Fortunately this is one of the brightest
clusters in our sample and we were still able to produce acceptable
quality images and spectra.
 
Figure~\ref{fig:image_ii}(c) shows the image
of the RXJ1701.3+6414 field. Two nearby sources are marked on the Figure; a QSO
(HS1700+6416) and a known cluster (A2246).  

The XMM-Newton data indicate an anomalously low r$_{c}$ value
in the $\beta$-fit, and this is supported by CHANDRA data
(\cite{Vikh_hiz}) where this cluster was observed with a
suspected cooling flow central brightness enhancement.  We therefore excluded the central bins (120 $h_{50}^{-1}$kpc)
and fixed the  r$_{c}$ to the CHANDRA value of 0.5 arcmin (204kpc).  Figure~\ref{fig:sb_ii} shows the
corresponding radial surface brightness distribution and resulting best fit
$\beta$ model; $\beta=0.64 \pm{0.05}$, $\theta_c=30$
arcseconds. This
value for $\beta$ is consistent with the CHANDRA measurement
  (0.62 $\pm$ 0.03). 

\begin{figure} 
\begin{center} 
\resizebox{\hsize}{!}{\includegraphics[angle=270,scale=0.5]{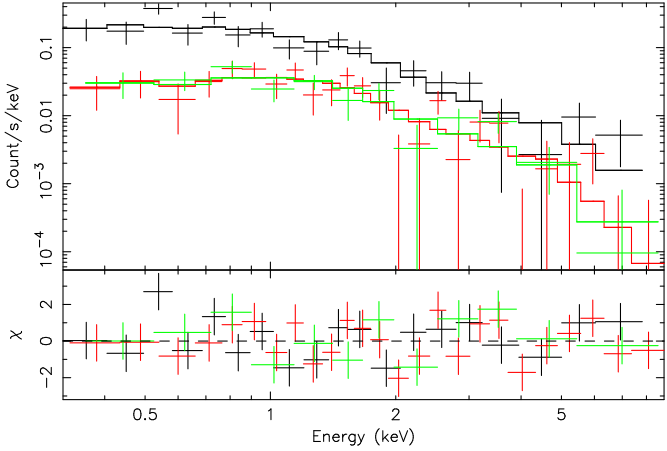}} 
\caption{\label{fig:rxj1701spec}  Spectral fit and residuals for
 RXJ1701.3; black - PN, dark  \& light grey - MOS} 

\end{center} 
\end{figure}

From a spectral extraction region with a radius of $120''$, we measured the 
following temperature, metal abundance and redshift (without core excision); 
$T_{\rm x}=4.5^{+1.5}_{-1.0}$~keV, 
$Z=0.24 \pm 0.2$, $z=0.44\pm0.02$ (Figure~\ref{fig:rxj1701spec}). 
The overall $\chi_{\nu}$ of the spectral fit was 58 / 53
degrees of freedom.  The hydrogen column density was 
fixed at the Galactic value (\nh$=2.6\times 10^{20}$ atoms 
cm$^{-2}$), and as a free parameter, its best 
fit value was very similar (\nh$=2.6\pm 1.5 \times 10^{20}$ atoms 
cm$^{-2}$). 

After excluding the 120kpc cooling flow region we obtain $T_{\rm
  x}=4.8^{+1.9}_{-1.3}$~keV.  The derived values for the bolometric luminosity and 
absorbed flux inside the viral radius are given in 
Table~\ref{tab:L}.  We were unable to apply the double subtraction 
technique to these data because of the enhanced proton background 
during the observation (see Figure~\ref{fig:protvig}). The poorer
  quality of data in this observation limits the quality of determination of T
  and L. Agreement with the CHANDRA data is acceptable for the
  temperature, but not the luminosity (5.8$\pm$0.5 keV and
  15.7$\times 10^{44}$erg/s respectively). 
 
\subsubsection{Abell 2246} 
 For completeness we also offer our interpretation of the Abell 2246
spectrum. From a spectral extraction region with a radius of $90 \arcsec$, and
using the in-field background subtraction technique, we measured the
following temperature, metal abundance and redshift by fitting to the
PN, MOS1 and MOS2 data simultaneously; $T_{\rm x}=2.7 ^{+0.6}_{-0.5}$
keV, $Z=0.32\pm0.13$, $z=0.22\pm0.04$. We note the consistency between
the redshift measured from the X-ray spectrum with the optically
determined value ($z=0.225$; \cite{struble}). Within a radius
$\sim$415 $h_{50}^{-1}$ kpc, 
the flux and bolometric luminosity are
3.4$\pm$0.2~10$^{-13}$ \ergcms (0.5-2keV) and 
%luminosity L$_{bol}\sim$
2.1~10$^{44}$\ergs respectively. Despite the rather restricted radius from which the
luminosity is determined, the local \LxTx  relation would predict
a slightly lower luminosity than this estimate.

\subsection{RXJ1354.2--0222} 
 
The RXJ1354.2--0222 field is shown in Figure~\ref{fig:image_ii}(d).  
A point source that was removed during analysis is
circled and its position listed in
Table~\ref{tab:srclist}. Figure~\ref{fig:sb_ii} shows the
corresponding radial surface brightness distribution and best fit
$\beta$ model that gives $\beta$=0.68 $\pm$0.06 and
r$_{c}$=33.6 $^{+4.9}_{-3.6}$ arcseconds.  ($r_c=248 ^{+36}_{-26}
h^{-1}_{50}$ kpc, Table~\ref{tab:L}).
 
\begin{figure} 
\begin{center} 
\resizebox{\hsize}{!}{\includegraphics[angle=270,scale=0.5]{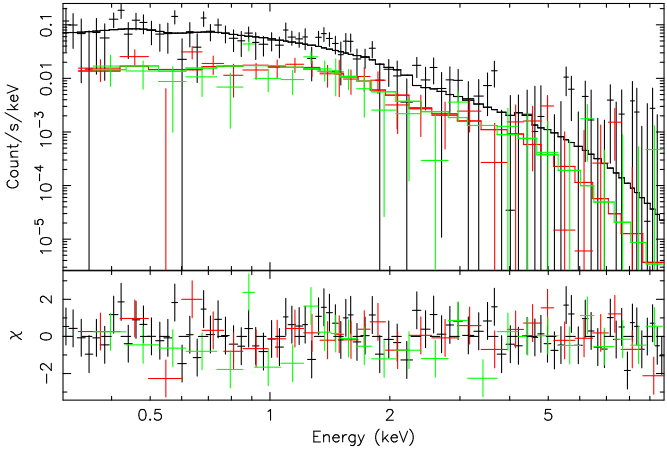}} 
\caption{\label{fig:rxj1354spec} Spectral fit and residuals for
 RXJ1354.2 black - PN, dark \& light grey - MOS} 
\end{center} 
\end{figure}

From a spectral extraction region with a radius of $120''$, and using
the in-field background subtraction technique, we measured; 
$T_{\rm x}=3.66^{+0.6}_{-0.5}$~keV, $Z=0.25
\pm 0.14$, (Figure~\ref{fig:rxj1354spec}).  The overall $\chi_{\nu}$ of
the spectral fit was 120 / 147 degrees of freedom. The fixed Galactic
hydrogen column value was \nh$=3.4\times 10^{20}$ atoms cm$^{-2}$,
but
\nh$=3.2\pm 1.2 \times 10^{20}$ atoms cm$^{-2}$) when free.  The
redshift  best fit value was $0.53
\pm0.04$.  
Similarly to RXJ1701.3, we were unable to apply the double-background
subtraction technique to these data because of the enhanced proton
background (see Figure~\ref{fig:protvig}).
 
\section{Discussion} 

\subsection{Luminosity-Temperature Relations} \label{sec:LT}

\begin{figure}
\begin{center}
\resizebox{\hsize}{!}{\includegraphics{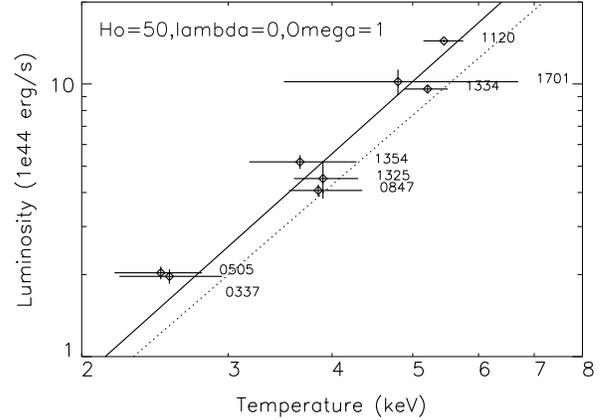}}
\caption{\label{fig:lxtx} $L_{\rm x}-T_{\rm x}$ relation for an EdS model ($q_0=0.5$)
and $H_0$=50 ~km~s$^{-1}$~Mpc$^{-1}$.  The solid line is the best fit
for our sample after cooling flow excision, the dotted line is the relation from Markevitch (1998)
for low red-shift clusters.}
%??watch out for typo in x-title (Temeprature) and maybe change ``L_bol''
\end{center}
\end{figure}

Table~\ref{tab:L} summarizes our analysis of the eight cluster
observations described above. The mean elemental abundance ($Z=0.28\pm
0.08$), core radius ($r_c=195\pm50 h_{50}^{-1}$ kpc) and $\beta$ (
$\beta=0.70\pm0.05$), are consistent with typical values measured at low
redshift (e.g. \cite{fukazawa};\cite{jonesforman};\cite{Mohr}). 
The clusters were
derived from well understood surveys (SHARC and 160SD) and thus should
be representative of the cluster population as a whole at these
redshifts. We have used our observations to determine the $L_{\rm
x}-T_{\rm x}$ relation at $0.45<z<0.62$ and to investigate
evolutionary effects. We plot luminosity versus temperature, with one
sigma errors, for the 8 cluster targets in Figure~\ref{fig:lxtx},
using
the values T$^{cf}$, L$^{cf}$ of Table 5. The
errors are clearly dominated by those of the temperature
measurements. 

Characterizing the $L_{\rm x}-T_{\rm x}$ relation as
\begin{center}
\begin{equation}
L_{\rm x} = L_{6} \left(\frac{T}{6 keV} \right)^{\alpha} %(1+z)^{A} 
\end{equation}
\end{center}
%??from this initial best fit, don't we get a $A$ value out too?
%?? are my explanations for d_L correct?
%?? is this ratio actually needed? if not remove from abstract also
and assuming an EdS cosmology, we find $\alpha$=2.7 $\pm$0.4, and
$L_{6}=15.9 ^{+7.9}_{-5.5} \times 10^{44}$\ergs~ using the bisector
variant of the BCES fitting package (\cite{BCES}). This relation is
shown as the solid line on Figure~\ref{fig:lxtx}. 
%Here
%$\frac{d_{L(z)}}{d_{L(z,EdS)}}^{2}$ refers to the ratio of the
%luminosity distance in the chosen cosmology compared to that in an EdS
%Universe. 
The slope of the relation is similar to most previous
measurements, e.g. for an EdS cosmology, $\alpha=2.64\pm0.27$
(\cite{markevitch}); $\alpha=2.33\pm0.43$ (\cite{allen98});
$\alpha=2.88\pm0.15$ (\cite{MonGus}); $\alpha=2.47\pm0.14$
(\cite{Ikebe}); $\alpha=2.82\pm0.32$ (\cite{novicki}).

%?? quoted errors
%\cite{markevitch}: 90%
%\cite{allen98}: 1 sdeviation
%\cite{MonGus}: 1 sigma
%\cite{Ikebe}: not stated
%\cite{novicki}: 68% on 1 interesting param
%\cite{Vikh_hiz}: 90% on 1 interesting param

\begin{figure}
\begin{center}
\resizebox{\hsize}{!}{\includegraphics{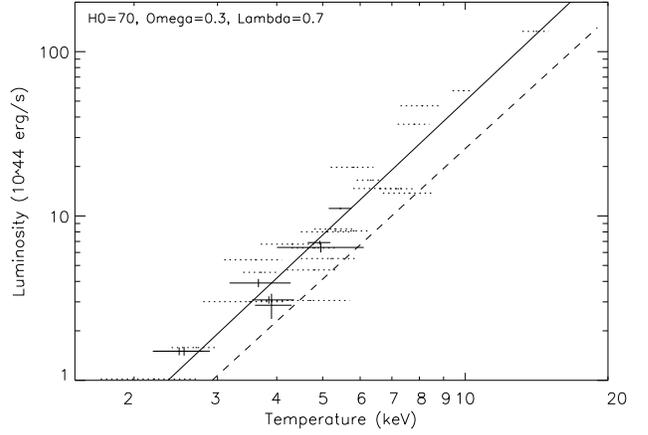}}
\caption{$L_{\rm x}-T_{\rm x}$ relation for $H_{o}$=70
$\OmL=0.7, \OmM$=0.3. The dotted horizontal lines are 
the CHANDRA sample reported by \cite{Vikh_hiz}, 
and the dashed diagonal 
line the Markevitch (1998)  \LxTx relation corrected to this
cosmology. The  solid crosses indicate the results from the clusters
in our study, corrected for this cosmological model, and the diagonal solid line 
is the best fit for our data.}

\label{fig:lxtx2}
\end{center}
\end{figure}

To characterize the possible evolution with $z$, we assume
\begin{center}
\begin{equation}
L_{\rm x} = L_{6}\left(\frac{T}{6 keV} \right)^{\alpha} (1+z)^{A }
\end{equation}
\end{center}
Adopting the low redshift $L_{\rm x}-T_{\rm x}$ relation from Markevitch
(1998) 
($\alpha=2.64\pm0.27$, $L_{6}=12.44\pm 1.08 \times 10^{44}$\ergs)
 we performed a
$\chi^{2}$ minimization on 
our data points corrected by a $(1+z)^{A }$ term.  The
Markevitch (1998) relation was derived from ROSAT and ASCA
observations of 30 clusters at $0.04<z<0.1$ ($\bar z=0.05$) and is plotted
as a dotted line in Figure~\ref{fig:lxtx}.  On average, the XMM-Newton data
points lie away from the Markevitch (1998) relation, suggesting an
evolutionary effect whereby clusters of the same temperature were more
luminous in the past.  The best fit value for $A $ is
0.68$\pm$0.26 ($\chi_{red}$=1.2). We have also determined the BCES best fit for our
sample when all the points are corrected with this evolution to a single redshift
(low redshift $z=0.05$ of Markevitch) bin, and find $\alpha=2.71\pm$0.22, and
$L_{6}=12.5^{+4.9}_{-3.5} \times 10^{44}$\ergs. We note that we have
chosen to compare with Markevitch (1998), rather than with the recent
study of 82 clusters with ASCA temperatures by Ikebe et al. (2002), because
the  Markevitch (1998) method to derive cluster temperatures is closer
to our own (Section~\ref{sec:MarkVikh}), thus we try to minimise any
systematic
effects from different data treatment. In Figure~\ref{fig:lxtx2} we
re-plot our data and the Markevitch (1998) relation after converting
them to a common cosmology of $H_0=70$ ~km~s$^{-1}$~Mpc$^{-1},
\OmL=0.7, \OmM=0.3$. In this cosmology we find
$A $=1.52$^{+0.26}_{-0.27}$.

For comparison, we include the data points obtained by Vikhlinin et al. 
(2002) from a compilation of Chandra observations of high-$z$ clusters
($0.39<z<1.26$) on Figure~\ref{fig:lxtx2} (after adjusting to $H_0=70$
~km~s$^{-1}$~Mpc$^{-1}$).  From this Figure, it is clear that both the
XMM-Newton and Chandra points lie away from the Markevitch (1998) 
relation and that there is no obvious systematic offset between the
XMM and Chandra data
(see Section~\ref{sec:MarkVikh}). Vikhlinin et al. 
(2002) also used
Markevitch (1998) as a low redshift benchmark to investigate $L_{\rm
x}-T_{\rm x}$ evolution. From their data, Vikhlinin et al. 
(2002) found $A=0.6\pm0.3$ for $\OmL$=0, $\OmM$=1 and
$A=1.5\pm0.3$ for a $\OmL=0.7, \OmM=0.3$ cosmology. 

%??quote all the fitted values for Lambda cosmology, also in abstract

%??did we include the Vikhlinin points in the Lambda-cosmo fit?

%??can we re-run their fit to confirm we get the same
%result from the same data?

%??i'm worried about the Ho conversion

The value of $A $ has important implications for our understanding
of structure formation and cluster evolution. In a self-similar model 
(Kaiser, 1981) cluster X-ray properties are driven by gravitational 
processes, such as shock heating, and a value of A$\sim$1.5 would be 
expected in most cosmological models.  A self-similar EdS model predicts 
exactly 1.5, and the value is only slightly different in a low density 
concordance model, the correction due to the redshift dependence of the 
virial density (\cite{bn})  being small. 
In a model where the cluster X-ray properties are influenced by 
non-gravitational processes, such as energy injection by AGN's or
supernovae, lower values of $A $ are predicted (e.g. \cite{tozzi}).  
The evolution deduced from our observations assuming an EdS 
model is significantly below the predicted value.  Adopting a low density 
concordance model, on the other hand, leads to a value consistent with 
predictions. We conclude that there is now evidence
from both XMM-Newton and Chandra for an evolutionary trend in the
$L_{\rm x}-T_{\rm x}$ relation.

Previous studies %have claimed an absence 
of $L_{\rm x}-T_{\rm x}$
evolution based on either ASCA (e.g. \cite{Mush97}; \cite{allen98};
\cite{novicki}; Sadat et al., 1998), ROSAT PSPC (e.g. \cite{Fairley}) 
or Chandra
(e.g. \cite{borgani01}; \cite{holden}) temperature measurements
found results consistent with no evolution (\cite{sbo}). When these studies quote values
for $A$, these values are generally smaller than those measured by us or
Vikhlinin et al. (2002), but are usually still at least one sigma
away from $A =0$. 
Sadat et al. (1998) 
found a positive evolution $A =0.5\pm 0.3$ for $\OmM=1,\OmL=0$
 and Novicki, Sornig, \& Henry (2002) 
$A =1.1\pm1.1$ for $\OmM=1,\OmL=0$, and $A=2.1\pm1.0$ for
$\OmL$=0.7, $\OmM$=0.3 respectively.  It is clear that we are  beginning to
probe evidence that  $L_{\rm x}-T_{\rm x}$ evolution, although 
many more clusters need to be
studied, and systematic biases (see below) examined in detail, before
solid conclusions regarding structure formation models can be drawn.

\subsection{Data Treatment and Systematics}
\begin{figure*} 
\begin{center} 
\begin{tabular}{cc} 
\includegraphics[scale=0.5]{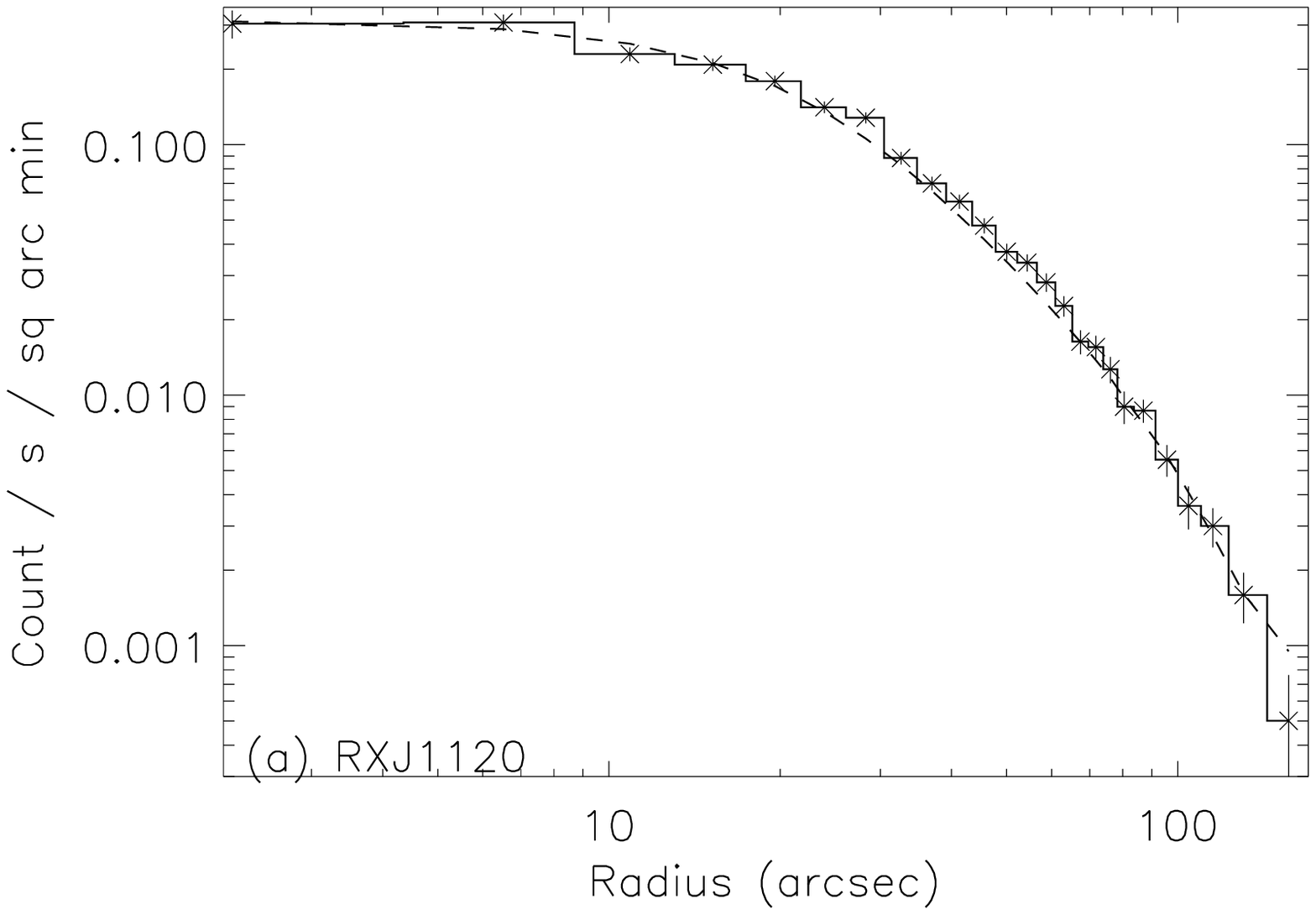}&\includegraphics[scale=0.5]{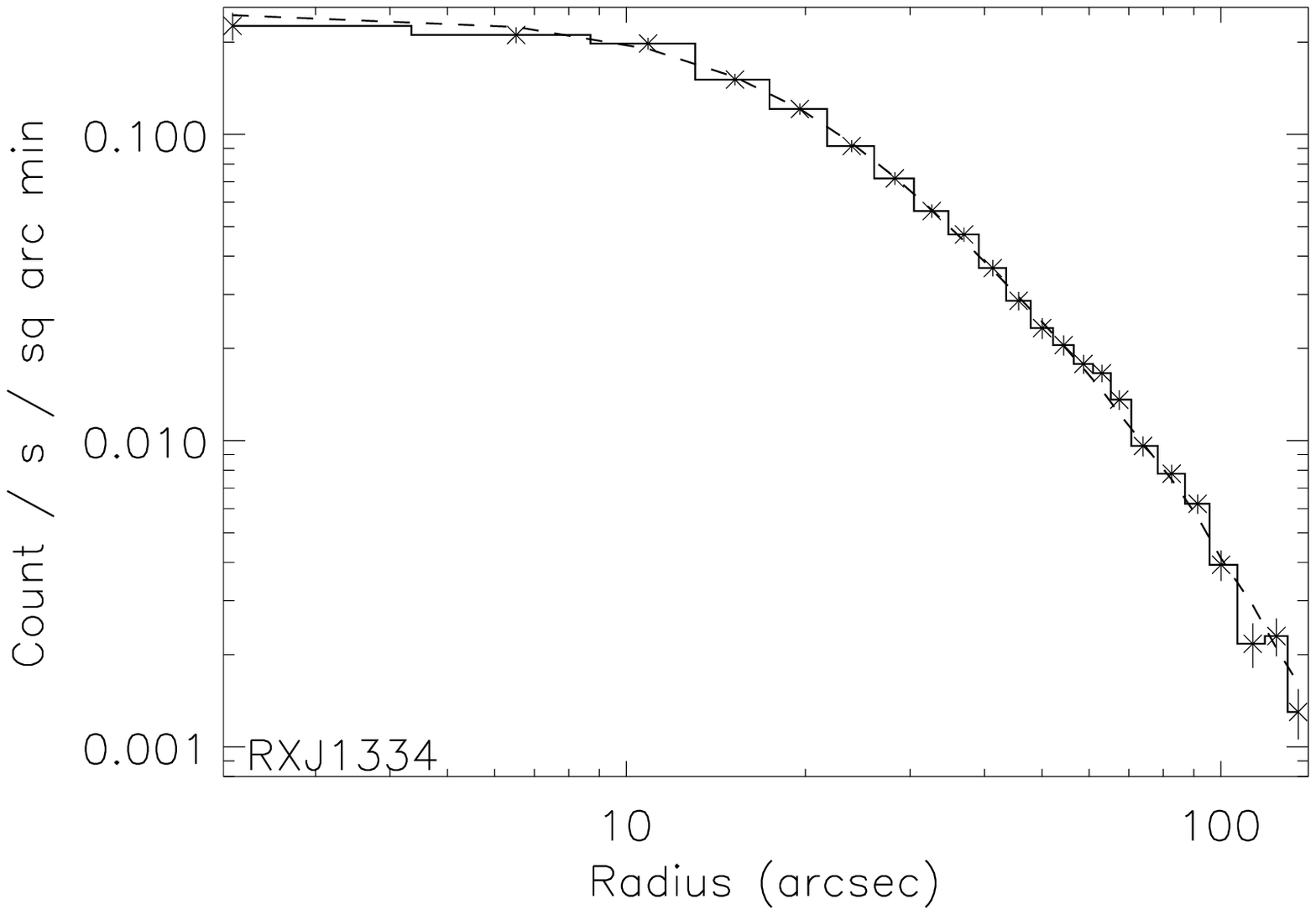}\\ 
\includegraphics[scale=0.5]{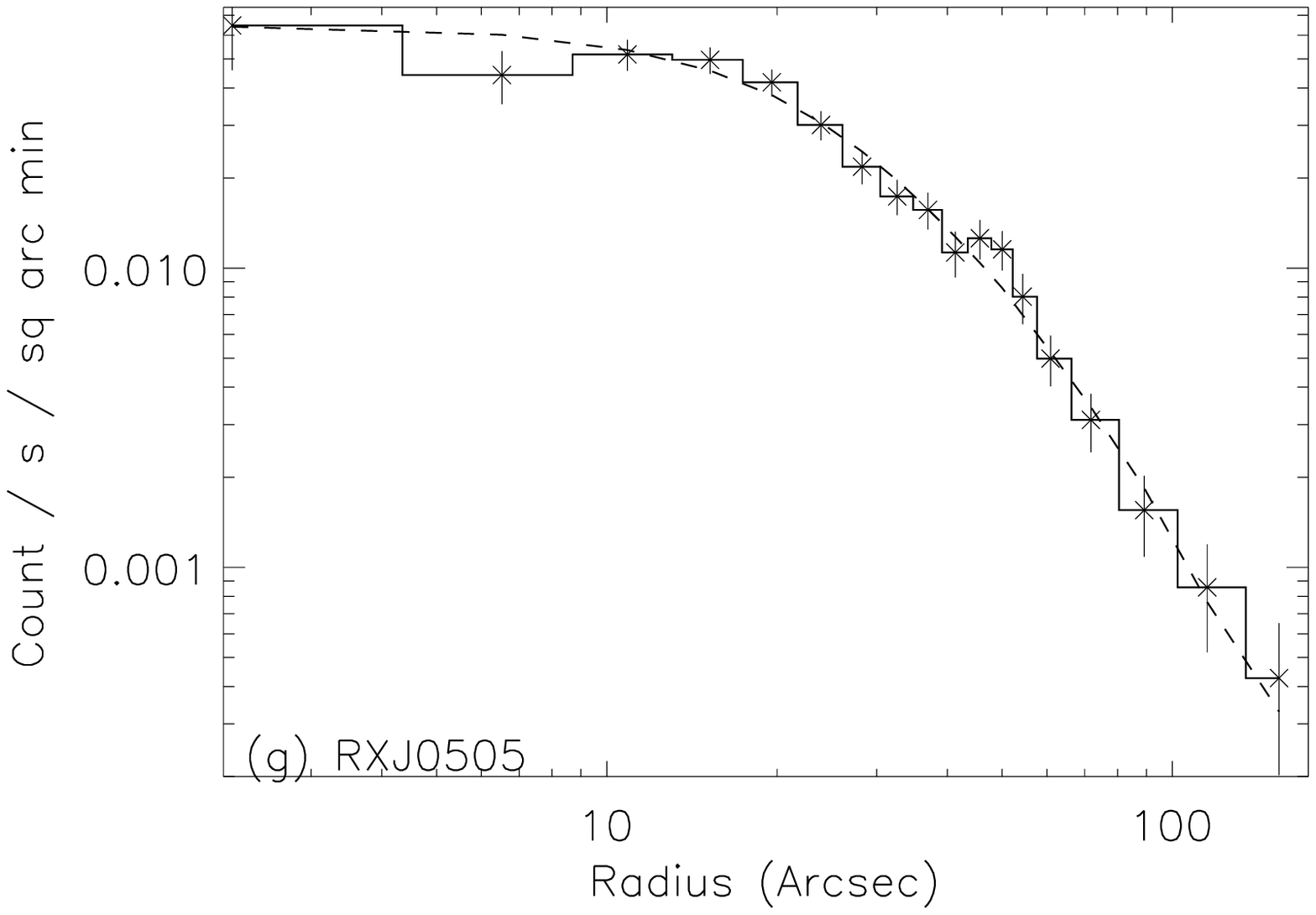}&\includegraphics[scale=0.5]{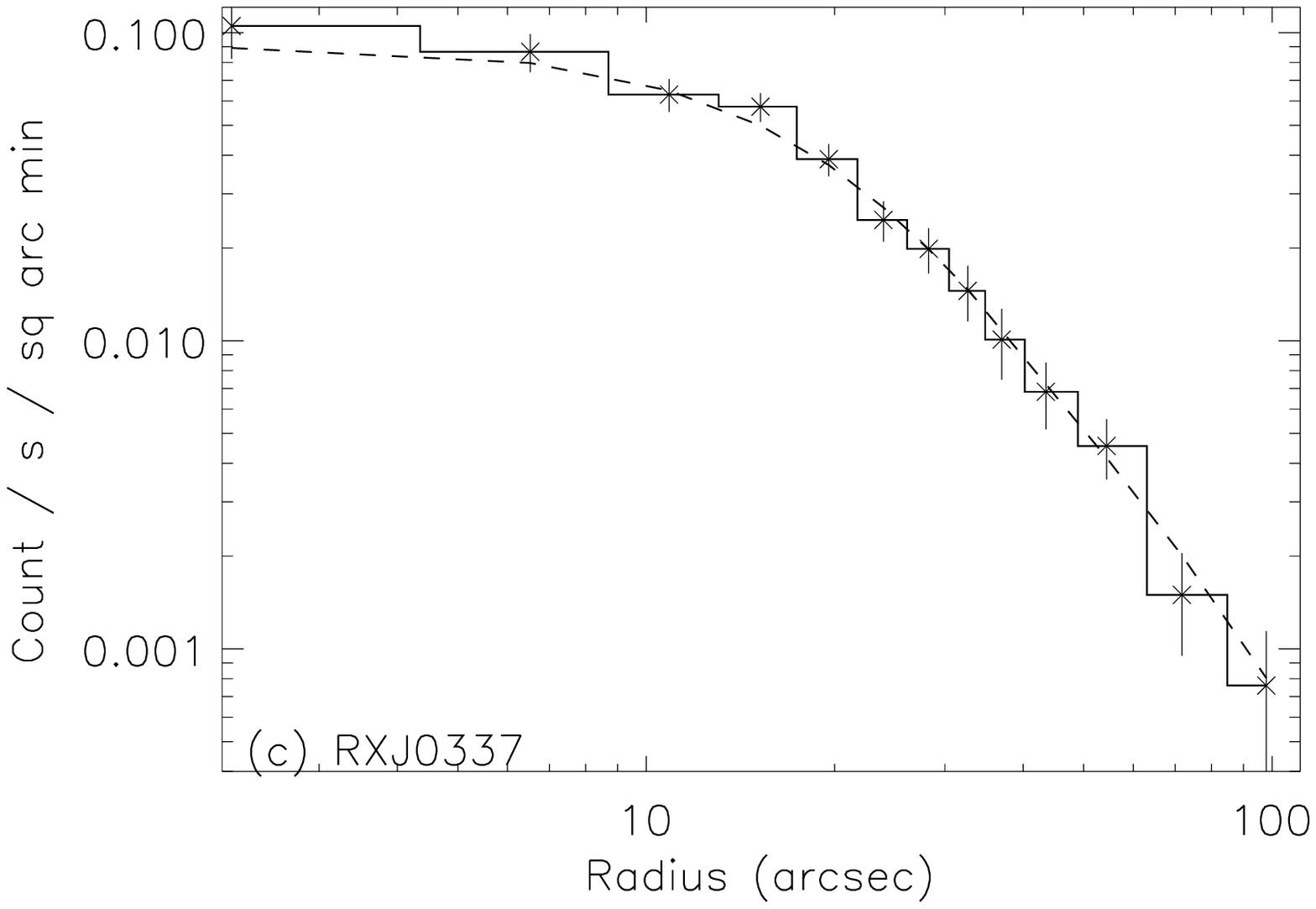}\\ 
\end{tabular} 
\caption{\label{fig:sb_i}  Surface brightness distribution of clusters, compared with
best $\beta$--model fit (dotted line). Reading clockwise from top
left: RXJ1120.1, RXJ1334.3, RXJ0337.7 and RXJ0505.3.}

\end{center} 
\end{figure*}  

We have made every effort to ensure that the $L_{\rm x}-T_{\rm x}$ relation
presented above is robust. We have adopted a uniform approach to the
spatial and spectral analysis of the eight clusters in the sample. We
have excised regions with clear signs of point source contamination
before extracting spectra and surface brightness profiles. We have not
used $\beta$-models to calculate total cluster count rates, but have
rather summed up the counts within a viral radius. We have tried to
ensure that our approach to background subtraction does not bias the
measured cluster parameters (see below).  We have used measured
temperatures to make conversions between count rates and fluxes and,
where possible, we have used data from all three EPIC cameras to
derive $L_{\rm x}$ and $T_{\rm x}$ values. For RXJ0847.2 and
RXJ1325.5, data were not available from all three cameras. However, we
are confident that the derived quantities for these two clusters are
robust, as we have shown, using the RXJ1120.1 observation, that
PN-only or MOS-only spectra are in excellent agreement.

The background subtraction has been complicated by the extended nature
of the targets and the high proton flare contamination in some of the
observations. We used  background template files
during the spatial analysis. These files covered a limited energy
range (0.3--4.5~keV), to minimize any calibration uncertainties. They
were filtered using the same rate filtering criteria as their
respective cluster observation and were normalized to it using the
particle background rates in the CCD regions outside the FOV. For
various reasons (see Section~\ref{sec:back}), we did not use
the background templates for the spectral analysis. Instead, we
adopted an in-field background subtraction technique. With this method
comes the concern that the proton background can be over-vignetted. We
have checked for this by applying a different method, the double
subtraction technique, to the RXJ1120.1 and RXJ1334.3 observations. We
found that the results from both techniques agree within the
statistical errors. For the two observations with the worst proton
flare contamination, RXJ1354.2 and RXJ1701.3, we applied a 2\% scaling
to the in-field background spectrum to compensate for potential
over-vignetting of the proton background.  In Figure~\ref{fig:protvig}
we demonstrate that this over-vignetting should not present a problem
for the other clusters in our sample.
 
Despite all these quality controls, we still cannot rule out the
possibility of there being some systematic bias in our results.  We
discuss possible sources of such bias and the potential impact on our
claimed $L_{\rm x}-T_{\rm x}$ evolution below.

\subsubsection{Impact of Cooling Core Clusters}
\label{sec:cflows}
%??we must put full L-T fit parameters in here
\begin{figure*} 
\begin{center} 
\begin{tabular}{cc} 
\includegraphics[scale=0.5]{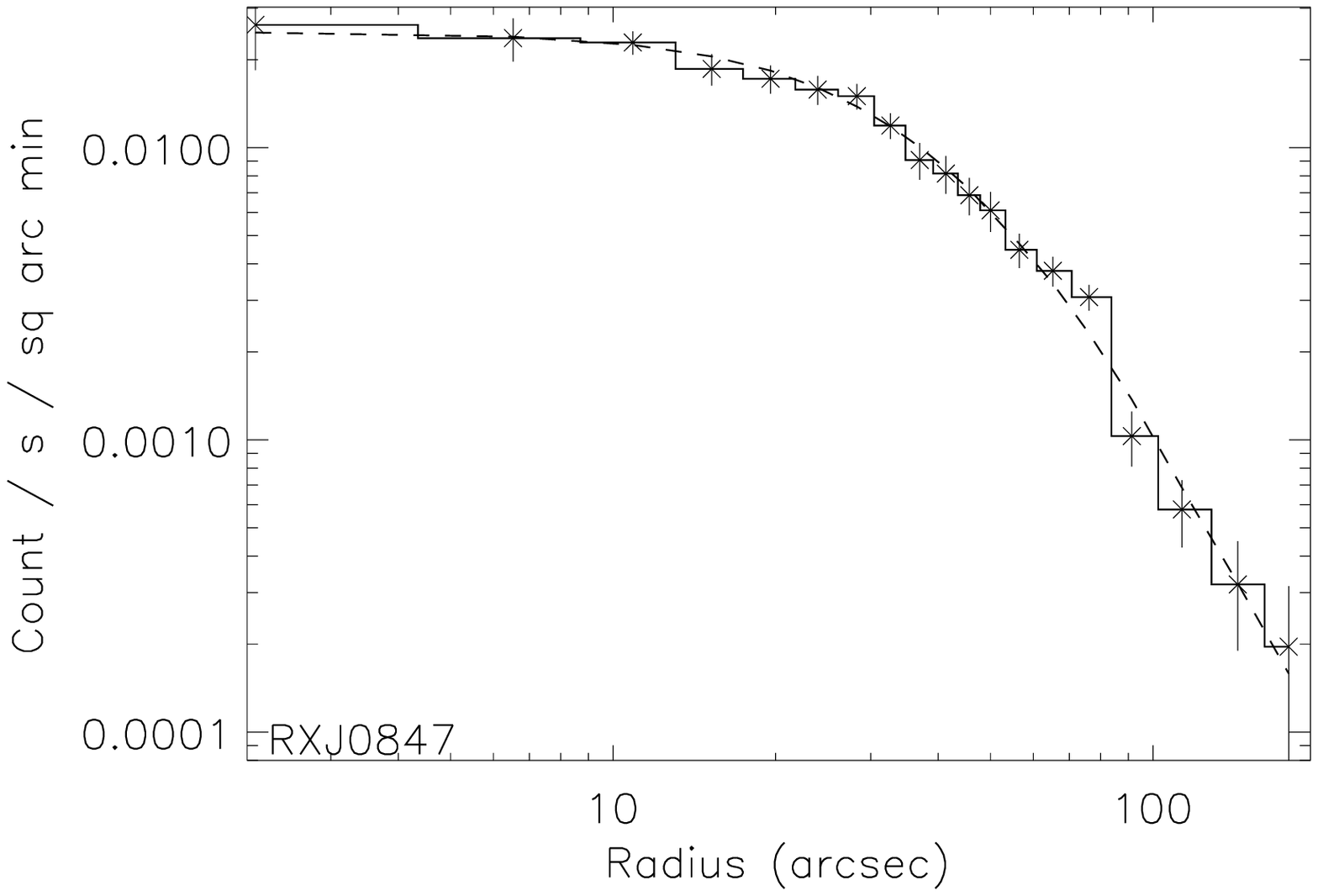}&\includegraphics[scale=0.5]{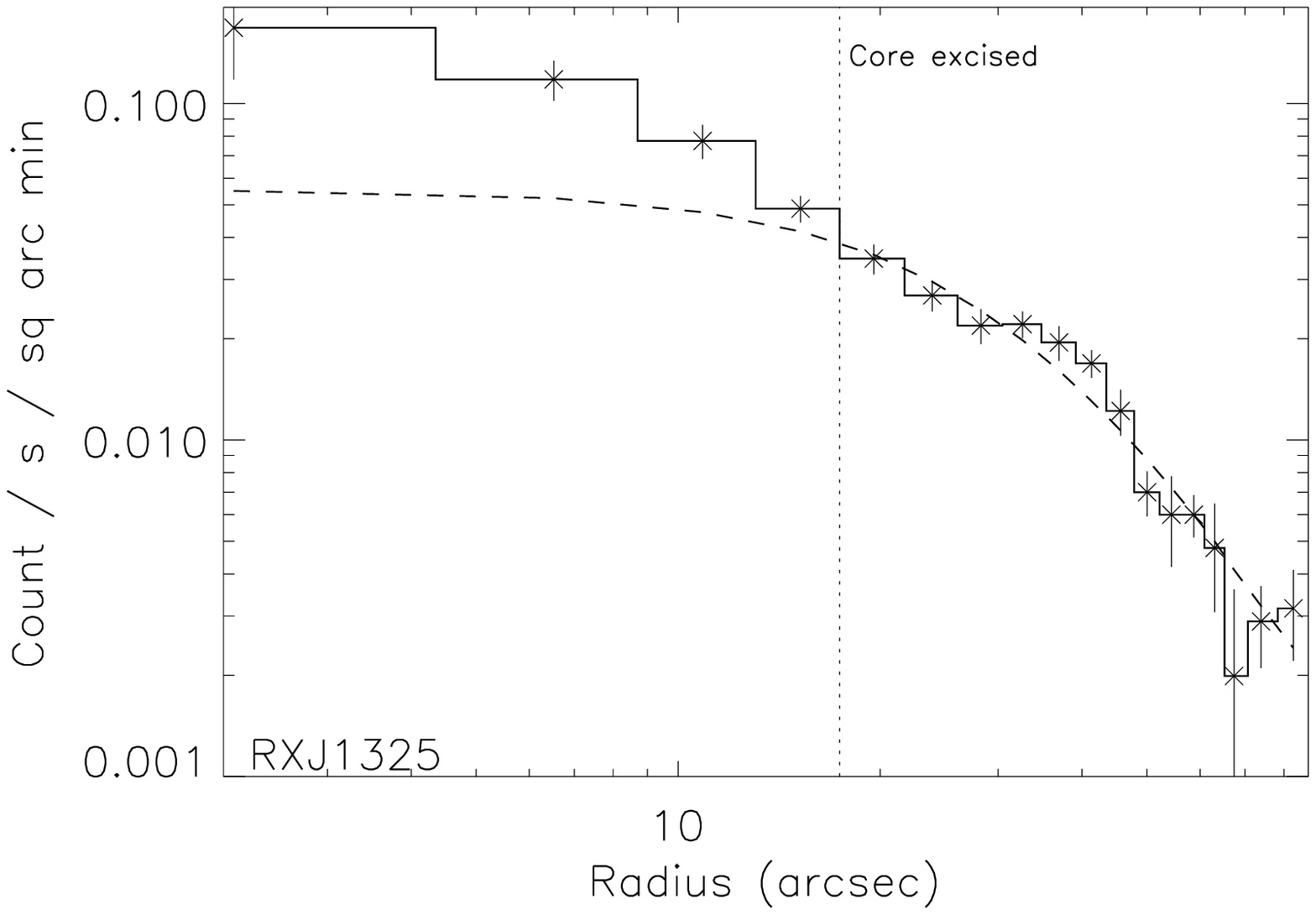}\\ 
\includegraphics[scale=0.5]{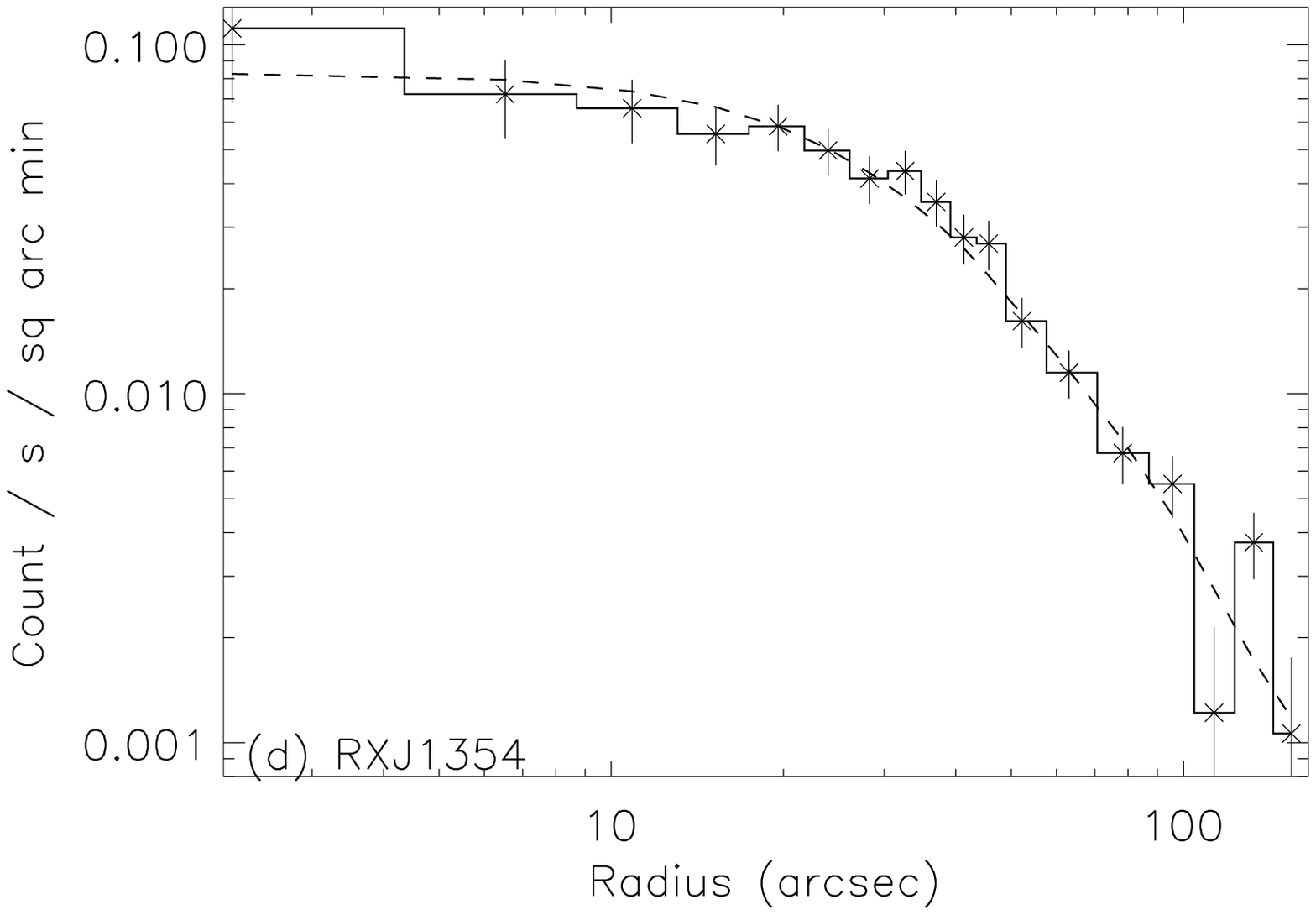}&\includegraphics[scale=0.5]{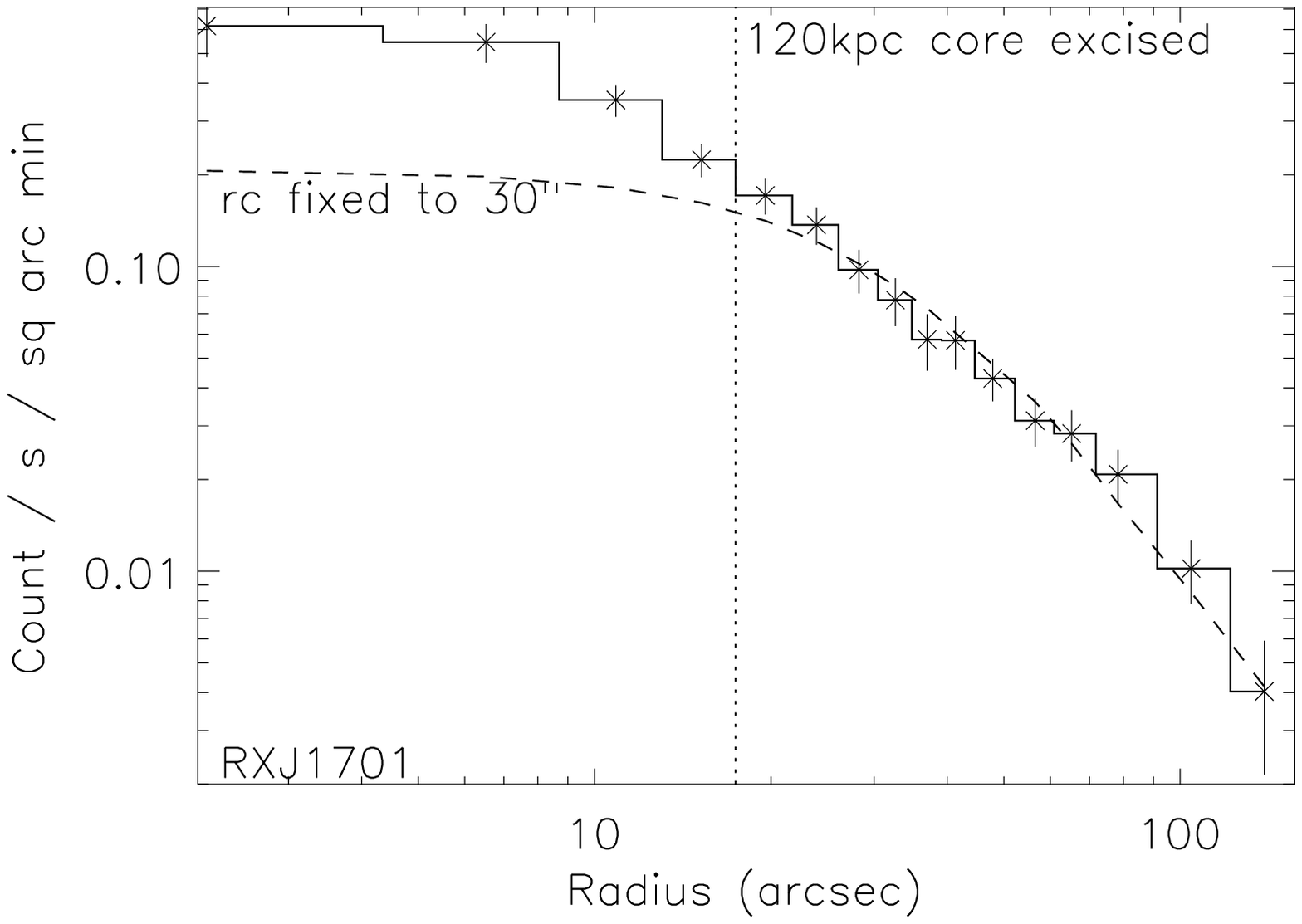}\\ 
\end{tabular} 
\caption{\label{fig:sb_ii}  Surface brightness distribution of clusters, compared with
best $\beta$--model fit (dotted line). Reading clockwise from top
left: , RXJ0847.2, RXJ1325.5, RXJ1701.1 and RXJ1354.2.}
\end{center} 
\end{figure*} 

Cooling  cores have the joint effect of boosting
the cluster luminosity and of reducing the measured global cluster
temperature. These effects have been shown to introduce scatter in the
$L_{\rm x}-T_{\rm x}$ relation and to modify its slope
(\cite{markevitch}; \cite{allen98}). Various approaches have been taken to alleviate
the effects of cooling cores on measured $L_{\rm x}-T_{\rm x}$
relations, e.g. by excluding cooling core clusters from the analysis
,\cite{MonGus}, by fitting two temperature models (\cite{Ikebe}), by
masking the central regions from obvious cooling core clusters
(\cite{Vikh_hiz}), or by masking the central regions from all clusters
(\cite{markevitch}). Based on the images (Figures~\ref{fig:image_i} \&
\ref{fig:image_ii}) and radial profiles (Figures~\ref{fig:sb_i} \&
\ref{fig:sb_ii} ) of the eight clusters in our sample, we do not
expect cooling cores to have much, if any, impact on the measured
$L_{\rm x}$ and $T_{\rm x}$ values; only RXJ1701.3 and RXJ1325.5 show
evidence of a central surface brightness excess. We have investigated
this by comparing the results before and after excising a circular
region \footnote {The radius of the excised region was 50 $h_{50}^{-1}$ kpc
except in the case of RXJ1701.3 and RXJ1325 where $\sim$120kpc was used, see sections
3.6 and 3.7} from the cluster cores. 

The results of the re-analysis can be found in
Table~\ref{tab:L}, (rows 4 and 5).  Temperatures changed very little -- always within the
statistical error -- after we excised the central $r=50 h_{50}^{-1}$
kpc region.
In two cases,
( RXJ1120.1 and  RXJ1701.3), the luminosity changes by
more than $1\sigma$, but still by only 5\% and 9\% respectively.
Without the excision the best fit \LxTx relation is $L_{6}=17.6 ^{+5.3}_{4.1}\times
10^{44}$\ergs~ and $\alpha$=2.66$\pm$0.25 for an $\OmL=0.0, \OmM=1.0$,
$H_0=50$ ~km~s$^{-1}$~Mpc$^{-1}$ cosmology 
at a typical redshift $z \sim 0.55$.  In
summary, we find the $L_{\rm x}-T_{\rm x}$ relations before and after
the core excision are consistent within the errors. However, we
caution that some residual cooling flow contamination may still remain
after this excision; $50 h_{50}^{-1} kpc \simeq 14''$, corresponding to an
encircled energy fraction of a little under 50\% of the on-axis mirror
PSFs.

\subsubsection{Incompleteness and Flux Errors in the ROSAT Catalogs}
\label{sec:sharc-flux}

Our $L_{\rm x}-T_{\rm x}$ evolution result implies that high redshift
clusters in our sample are more luminous than clusters of the same
temperature at lower redshift.  We interpret this as evidence for a
general evolutionary trend in the \LxTx relation, but
it might also reflect an underlying selection bias in the cluster
catalogs from which the sample was drawn, e.g. if the SHARC and 160SD
surveys were biased toward merging systems. It is possible  that the
luminosity and temperature of clusters can change dramatically during
the process of a major merging event (\cite{ricker}).  Based on the
XMM-Newton images of the 8 clusters studied here
(Figures~\ref{fig:image_i} \& \ref{fig:image_ii}), such a bias does
not appear to exist; only RXJ1354 has evidence for a possible
  double component, and excluding this from the best fit $L_{\rm x}-T_{\rm x}$
relation changes the luminosity normalisation only $\sim$2\%. As long
as the angular resolution of observations allows sufficient
discrimination against obvious mergers, then presumably an ensemble
sample of local and distant clusters are similarly affected by merger
boosting of luminosity.  

Moreover, we find no evidence for any incompleteness
in the SHARC and 160SD surveys, at least in the small area covered by
these observations. No new clusters were found in the fields
surrounding the seven SHARC clusters in our sample.  In the field of
the 160SD cluster, RXJ0847.2, we have discovered a new X-ray cluster
(XMMU J084701.8+345117). However this cluster was too faint to have been
included in the original 160SD survey.  We estimate the ($r<r_v$) flux of this
object to be $1.3\times 10^{-14}$ \ergcms (based on a fitted
temperature of $T_x=1.8$ keV and an assumed redshift of $z=0.56$),
compared to the 160SD flux limit of $3.7\times 10^{-14}$ \ergcms, so
this cluster is too faint to have been included in the original 160SD
survey. In summary we do not find evidence for any intrinsic bias in
the SHARC or 160SD survey selection functions that could mimic $L_{\rm
x}-T_{\rm x}$ evolution.

Another possible way to mimic the claimed evolution would be to
systematically overestimate the $L_{\rm x}$ values. We have
investigated this by comparing our flux measurements to those
published in several works (\cite{brightsharc2}; \cite{southern}; \cite{Vikh}).
  We
report here that a systematic offset does appear to exist, but in the
opposite direction to that needed to mimic evolution.  The XMM-Newton
fluxes quoted in Table~\ref{tab:L} are all lower than previously
determined values, typically by 20\% but as much as by 40\% in the
case of RXJ0505.3.  We note that this discrepancy persists regardless
of the ROSAT catalog from which the cluster was selected. The average
discrepancy between the Southern SHARC (4 clusters), Bright SHARC (3
clusters) and 160SD (3 clusters) fluxes and the XMM-Newton fluxes is
22\%, 21\% and 16\% respectively.

Improvements in point source rejection, count rate to flux conversions
and the signal to noise have meant that the typical statistical error
on the XMM-Newton determined fluxes in Table~\ref{tab:L} is $\simeq
3\%$, compared to $\simeq 10\%$ for the ROSAT determinations.
Therefore, even accounting for a possible offset in the absolute cross
calibration of the two observatories, it is clear that a systematic
flux offset does exist.  There are likely to be several reasons for
this offset. Of these, the use of an incorrect spatial model, to
convert between an aperture flux and a total flux, is probably the
most significant. As shown in \cite{brightsharc2}, the use of a
generic ($\beta=0.67$, $r_c=250 h_{50}^{-1}$ kpc), rather than best
fit, $\beta$-model resulted in a typical enhancement of 10\% in the
total cluster flux. In our XMM-Newton analysis, we do not use the
$\beta$-model fits to determine the total flux. Instead we directly sum
up the counts within a virial radius. For some clusters this
difference in approach can explain the entire ROSAT to XMM-Newton
discrepancy. For example, in the case of RXJ1334.3, our estimate of
$r_v$ is almost identical to the $r_{80}$ radius within which 80\% of
the total flux from a generic $\beta$-model would
fall. \cite{brightsharc2} divided the ROSAT counts within $r_{80}$ by
0.8 to estimate the total count rate. For RXJ1334.3 this division
would artificially boost the flux by 25\% which more than accounts for
the 22\% the discrepancy between the XMM-Newton and Bright SHARC
values.

For the four Southern SHARC clusters in our sample, we have
recalculated the total ROSAT PSPC fluxes using the $\beta$ and $r_c$
values given in Table~\ref{tab:L} (in \cite{southern} generic values
were used; $\beta=0.67$, $r_c=250 h_{50}^{-1}$ kpc).  By doing so, we
improve the consistency between the ROSAT and XMM-Newton fluxes to
within 5\% for RXJ1354.2 and RXJ1325.5 and within 15\% for
RXJ0337.7. For RXJ0505.3 there remains a large, 35\%, discrepancy
which requires further investigation.

For the three Bright SHARC clusters in our sample, we have gone back
to the original \cite{brightsharc2} data and measured count rates
inside $r_v$. Doing so reduces the ROSAT determined flux by $\simeq
30\%$ in the case of both RXJ1120.1 and RXJ1334.3, i.e. so that they
are $\simeq10\%$ lower than the values quoted in Table~\ref{tab:L}. In
the case of RXJ1701.3, the measured ROSAT flux actually increased
slightly when measured inside $r_v$ rather than $r_{80}$. On closer
examination of the ROSAT data, it was apparent that the nearby QSO and
cluster (see Figure~\ref{fig:image_ii}) were contaminating both the
source and background apertures in the PSPC image. When the source and
background were accumulated only from the (source free) region to the
West of the cluster, the ROSAT flux within $r_v$ dropped to within
$\simeq10\%$ of the XMM-Newton determined value. These examples
demonstrate the importance of using XMM-Newton to re-calculate fluxes
and luminosities for high redshift clusters detected at low signal to
noise by ROSAT.  We stress that, despite these flux uncertainties, the
SHARC and 160SD catalogs are still fair representations of the
underlying cluster population and can still be used to probe the
$L_{\rm x}-T_{\rm x}$ relation.  It is also important to note that the
these flux uncertainties do not apply to the ROSAT observations used
by \cite{markevitch} to determine $L_{\rm x}$ values for the 30 low
redshift clusters in his $L_{\rm x}-T_{\rm x}$ analysis. Those
observations have exquisite signal to noise, so that the $L_{\rm x}$
values derived from them will be limited only by the absolute
calibration of the instrument.

 The fact that we find a \LxTx relation consistent with the 
low redshift value argues against a bias in our selection towards the
more luminous tail of the cluster population; for either 
\begin{itemize} 
\item all the clusters must be biased in the same way, or
\item they are biased in just such a way as to offset any real
  evolution in the value of the slope.
\end{itemize}

Nevertheless the reader might be cautioned that should there remain a common
  bias for all cluster samples, despite these cross-checks, it is
  possible the clusters selected represent the brightest portion of
  the intrinsic high-z sample, and thus remain on the high luminosity
  end of the true population, hence mimicking the luminosity evolution
  we see. Further validation of the ROSAT data with CHANDRA and
  XMM--Newton should help to close this issue.

\subsubsection{Comparison with Markevitch (1998) and Vikhlinin et al. (2002)}
\label{sec:MarkVikh}

We come to very much the same conclusions regarding \LxTx
 evolution as did \cite{Vikh_hiz}. Given that these conclusions 
provide evidence for significant evolution, contrary to previous analyses, 
 it is important to
investigate whether this concordance is genuine or coincidental.  Both
studies rely on \cite{markevitch} for the low redshift comparison, so
let us first consider how our XMM-Newton clusters might appear either
too luminous, or too cool, compared to the \cite{markevitch} sample:

\begin{enumerate}

%??check the r_v values again, I took them from an email from April 8th
%??are 6 of 8 larger than 1Mpc?
%%AB It depends slightly on the calibration of M-T (and cosmology....). 
%%AB From my current value they all are above 1 Mpc with 2 cluster very close to it.

%??we could calculate the Lx's inside 1Mpc to quantify the effect.

%%AB: I think the main point is that at zero redshift  1Mpc is significantly 
%%AB: less than rv. However taking the 4% of Vikhlinin is enough to conclude. 
%%AB: I would not include the comments on rv. Accordingly to scaling rules  
%%AB: one would like to compute Lx within 1Mpc/(1+z) on our cluster. To check 
%%AB: if this is what Vikhlinin is speaking of this or not.

\item

{ 

Markevitch 1998 measured cluster fluxes by summing the count rate
inside circular apertures. However, rather than using a cluster
specific virial radius, he used a fixed metric aperture of $r=1
h_{50}^{-1}$ Mpc and ignored any flux that might lie outside this
radius.  
%%AB Of the eight clusters in our sample, six had $r_v$ values
%%AB larger than $r=1 h_{50}^{-1}$ Mpc, and so their $L_{\rm x}$ values
%%AB will be higher than if we had used the Markevitch (1998)  method.
However, this would be a very small effect; Vikhlinin et al. (2002) estimate
that the no more than 4\% of the total flux would be missed by
adopting a fixed $r=1 h_{50}^{-1}$ Mpc aperture.

}

\item{ 

When Markevitch 1998 excised the central $r=50 h_{50}^{-1}$ kpc region from
all the clusters in his low redshift sample before measuring the flux,
he  applied a multiplication factor of 1.06 to account for flux
falling inside the excised region. We carried out an excision
 and flux lost correction  on our eight clusters with a
 cluster-specific
surface brightness model.
(Section~\ref{sec:cflows}, and Table~\ref{tab:L}), we found that for six
of the eight clusters, the measured $L_{\rm x}$ values dropped
slightly. The difference in
our techniques could systematically affect cluster brightness,
compared with  the low redshift counterparts, but only at the few percent level.

}

\item
{

%??I need to find a few examples for this

Uncertainties in the cross calibration between XMM-Newton and the
instruments used for the \cite{markevitch} analysis (ROSAT and ASCA)
may mimic evolution in the $L_{\rm x}-T_{\rm x}$ relation. At this
time it is not possible to rule out that possibility; to date, a full
comparison of XMM-Newton and ASCA determined cluster temperatures has
not been carried out.

 }

%careful tests on a large set of clusters observed at high signal to
%noise by multiple observatories needs to be undertaken first. T

\item 
{
%?? I am looking for an XMM to ROSAT/ASCA comparison at low z.

Ikebe et al., 2002 have used a different approach to \cite{markevitch}
to analyse ASCA observations of low redshift clusters. They measure
temperatures that are on average lower than those measured by
\cite{markevitch}, with the trend becoming more pronounced as $T_{\rm
x}$ increases. However, this should not impact our conclusions
regarding $L_{\rm x}-T_{\rm x}$ evolution, given that we used a
similar technique to \cite{markevitch} to measure $T_{\rm x}$.

%??someone should run a markevitch-style vs Ikebe-style fit to the
%same cluster using XMM data. Hopefully the ref won't ask us to do that, since 
%it'd be hard and our data aren't up to that.
}

%\item 
%{

%Residual proton contamination in our images could have affected the
%$T_{\rm x}$ values in Table~\ref{tab:L} in such a way so as to mimic
%$L_{\rm x}-T_{\rm x}$ evolution.  After vignetting correction, the
%proton background level is enhanced towards the edges of the FOV. Thus
%the use of an in-field background subtraction might result in an over
%subtraction of the proton background. This would artificially soften
%the net source spectrum, because the background spectrum is so hard.
%With this in mind, we have taken every precaution to mitigate the
%effects of the proton background on our $T_{\rm x}$ values
%(section~\ref{sec:back}).

%??does this repeat stuff from earlier too much

%}
\item
{
Some estimate of the effect of external systematic error {\em could}
be taken by considering the value of $A $ in fitting to a different
low-z \LxTx sample. We have done so using the \cite{MonGus} relation
and find $A=0.95 \pm 0.2$ (again in the Einstein de Sitter
cosmology). Therefore, despite a possible contamination in that
sample from cooling flows we still see evidence for evolution
}
\end{enumerate}

%Chandra can excise point sources better

%??add clusters with both Chandra and ASCA and Chandra and XMM (1ES, 0152)

We note that the first two concerns listed above do not
apply to the Vikhlinin et al., (2002) study, as they deliberately adopted the same
methodology as \cite{markevitch} and the last three concerns are not
likely to be important either; Vikhlinin et al., (2002) have carried out a
cross comparison of Chandra and ASCA $T_{\rm x}$ values for $\simeq
20$ low redshift clusters and find them to be in agreement at better
than the 5\% level, with no systematic offset. Despite this, we should
still consider the possibility that both studies (ours and
\cite{Vikh_hiz}) might be pointing to the wrong value of $A $. 
At present we cannot quantify how
factors such as uncertainties in the cross-calibration of Chandra and
XMM-Newton, especially at low energies, might impact our
conclusions. We note in particular that Vikhlinin et al., (2002) applied an
empirical factor of 0.93 to improve the cross-calibration of the ACIS
front- and back-illuminated CCDs below 1.8keV. Other differences
between the observatories may also be important. For example, it will
be easier to account for point source and cooling core contamination
using Chandra data, because of the improved spatial
resolution (see for example the case of RXJ1701.3). Alternatively, it should be easier to correctly account
for the various particle (cosmic ray and proton) backgrounds using
XMM-Newton data, because its CCD's cover a larger area both inside and
outside the FOV. In summary, although we cannot rule out the possibility that a
combination of factors have lead us to measure an artificially large
value for the $A $ parameter, we believe that our XMM measurements provide clear evidence for evolution in the $L_{\rm x}-T_{\rm x}$ relation.

%??add examples of clusters with both XMM and Chandra

%??add final final sentence

%?? not mentioned now:
%reduces the energy leverage and ability to determine absorption correction. 

%-----------------------------------------------------------
%?? add nH and Z values used during spectral fit to this table 
%?? make sure the correction to r_v factors are up to date 
%??are these absorbed or unabsorbed fluxes:  check section 3.7 also
\small
\begin{table*} 
\begin{center} 
\begin{tabular}{l l l l l l l l l}\hline\hline 
&RXJ0337.7&RXJ0505.3&RXJ0847.2&RXJ1120.1&RXJ1325.5&RXJ1334.3&RXJ1354.2&RXJ1701.3\\ \hline 
centroid RA & 03:37:45&05:05:19& 08:47:10&11:20:07&13:25:37&13:34:20&13:54:17&17:01:24\\ 
centroid Dec& -25:22:32&-28:48:50&+34:48:54&+43:18:04&-38:25:44&+50:31:02&-02:21:44&  +64:14:12 \\ \hline 
&&\multicolumn{5}{c}{Cooling Flow Excised}&&\\ \hline
$T_{x}^{cf}$ (keV)& 2.52 & 2.56 & 3.91 & 5.35 & 3.77 & 4.98 & 3.86 & 4.8\\  
& $^{+0.36}_{-0.32}$ & $\pm$0.3 & $^{+0.5}_{-0.35}$ & $^{+0.42}_{-.32}$ & $^{+0.4}_{-0.36}$ &$^{+0.26}_{-0.32}$ & $^{+0.62}_{-0.55}$ & $^{+1.9}_{-1.3}$\\ \hline 
$L_{bol}^{cf}$~10$^{44}$ & 1.91 & 1.97 & 3.89 & 13.7 & 4.2 & 9.47 & 5.3 & 10.2\\ 
 erg s$^{-1}$ & $\pm$0.17 & $\pm$0.16 & $\pm$0.7 & $\pm$0.23 &
 $\pm$0.8 & $\pm$0.24 & $\pm$0.4 & $\pm$1.1 \\\hline  
&&\multicolumn{5}{c}{No Cooling Flow Excision}&&\\ \hline
$T_{x}$ (keV)& 2.6 & 2.5 & 3.62 & 5.45 & 4.15 & 5.20 & 3.66 & 4.5\\  
& $\pm$ 0.35& $\pm$0.3 & $^{+0.58}_{-0.51}$ & $\pm$0.3 &$^{+0.4}_{-0.3}$ &  
$^{+0.26}_{-0.28}$ & $^{+0.6}_{-0.5}$ & $^{+1.5}_{-1.0}$\\ \hline 
$L_{bol}$~10$^{44}$ & 1.97 & 2.03 & 4.07 & 14.4 & 4.5 & 9.59 & 5.18 & 11.1\\ 
 erg s$^{-1}$ & $\pm$0.1 & $\pm$0.11 & $\pm$0.2 & $\pm$0.2 & $\pm$0.7 &  
$\pm$0.27 & $\pm$0.3 & $\pm$1.0 \\\hline 
Flux (0.5--2)& 4.37 & 5.64 & 7.04 & 24.5 & 8.0 &14.1 &9.8 &24 \\ 
10$^{-14}$cgs & $\pm$0.2 & $\pm$0.2 & $\pm$0.3 & $\pm$0.54 & $\pm 0.03 $ &  
$\pm$0.2 & $\pm$0.5 & $^{+1.5}_{-1.2}$\\ \hline
\nh (10$^{20}$ )&0.99&1.5&3.2&2.1&4.8&1.05&3.4&2.6 \\
(atom cm$^{-2}$)&&&&&&&& \\\hline 
$\beta$ & 0.76 & 0.66 & 0.81 & 0.77 & 0.64 & 0.66 & 0.68 & 0.64\\ 
& $^{+0.08}_{-0.04}$ &$^{+0.05}_{-0.04}$ &$\pm$0.045&$\pm$0.03&$^{+0.09}_{-0.07}$&$\pm$0.02&$\pm$0.055&$\pm 0.05$\\\hline 
r$_{c}$ (kpc)&145&164&307&209&115&154&248&204\\ 
h$_{50}^{-1}$&$\pm$18&$\pm$ 17&$\pm$30&$^{+9}_{-8}$&$\pm
20$&$\pm$10&$^{+36}_{-26}$&{\em fixed}\\\hline 
Abundance &0.38&0.17&0.30&0.47&0.31&0.15&0.25&0.24\\ 
(Z)&$\pm$0.09 &$\pm$0.08&$\pm$0.28&$\pm$0.09&$^{+0.19}_{-0.17}$ & $\pm$0.08  
& $\pm$0.14 & $\pm 0.2$\\\hline 
$z$ (optical) &0.577 &0.51&0.56&0.60&0.445&0.62&0.551&0.45\\ \hline 
Normaln.& 1.68 & 2.12 &2.65 &9.42 & 2.7 &5.61 &4.01 & 5.58 \\ 
mekal 10$^{-4}$& $\pm$0.10 & $\pm$0.08&$^{+0.12}_{-0.11}$ &$^{+0.23}_{-0.17}$  
&$^{+0.08}_{-0.07}$ &$\pm$0.16 &$^{+0.14}_{-0.23}$ &$^{+0.50}_{-0.40}$\\ \hline 
Radius ($''$)& 120 & 120 & 120 & 145 &90 & 120 & 120 & 120\\ 
Spect. extract.&&&&&&&&\\ \hline 
Fract$_{r}$ counts  & 0.90 & 0.87 & 0.79 & 0.965 & 0.52 & 0.84 & 0.83 & 0.81\\ 
within $r_v$&&&&&&&&\\ \hline

\end{tabular} 
\caption{\label{tab:L} Summary of cluster parameters for EdS model H$_{o}$=50, q$_{o}$=0.5.  
Spectral fitting errors, L,$\beta $ and r$_{c}$ are 1$\sigma$ on one 
parameter. Fluxes are the measured, absorbed fluxes in ROSAT 
band. Fract$_{r}$ is the fractional correction made from the spectral 
extraction region, to the total counts within the $r_{v}$ radius 
after accounting for point source excision, inter-chip gaps loss 
etc.. The 3rd \& 4th rows summarise the data for the case when the
core region has been excised from the
cluster core. (Except for RXJ1701.3, where following the CHANDRA data, we excise a cooling flow enhancement to 120kpc). } 
\end{center} 
\end{table*} 
\normalsize 
 
\section{Acknowledgments} 
We thank M Arnaud for very useful comments concerning the analysis and
interpretation of the data. We also thank M Markevitch and
A. Vikhlinin for their assistance.  M Watson is acknowledged for
allowing us to use data from the GT observation of IRAS
13224-3809. The EPIC instrument team are thanked for their continued
help in improving the calibration knowledge and help in understanding
various effects related to the instrument background. We thank
  the referee, F Castander for the careful review and comments that
  helped to improve the interpretations. DJB acknowledges
the support of NASA contract NAS8-39073 (CXC).  AKR and RCN
acknowledge support from the NASA-LTSA program, contract NAG5-11634
and the hospitality of the Durham University Physics department during
the summer of 2002. These data were obtained from observations made by
the {\em XMM-Newton} Observatory. This is an ESA science mission with
instruments and contributions directly funded by ESA Member States and
the USA (NASA). The XMM-Newton $\Omega$ Project GT data were provided
from a combination of data-rights holders, including the CDS
Strasbourg (XMM Science Survey Centre), CEA-Saclay (XMM EPIC PI Team)
and the XMM-Newton SOC.

\clearpage 

\onecolumn
\appendix 
\section{Excised Point Sources} 
\begin{table}[h] 
\begin{tabular}{l l l } \hline\hline 
XMM-Newton ID&RA&Dec\\ 
&2000&2000\\ \hline 
Field RXJ0337.7 -- 2522&&\\ 
XMMU J033737.8 -- 252318 & 03:37:37.8 &  -- 25:23:18.0\\ 
XMMU J033743.7 -- 252326 & 03:37:43.7 &  -- 25:23:26.7\\ 
XMMU J033742.9 -- 252208 & 03:37:42.9 &  -- 25:22:8.4\\ 
XMMU J033745.9 -- 252206 & 03:37:45.9 &  -- 25:22:6.2\\ 
XMMU J033747.2 -- 252214 & 03:37:47.2 &  -- 25:22:14.9\\ \hline 
Field RXJ0505.3 -- 2849&&\\  
XMMU J050512.5 -- 285034 &05:05:12.5&      -- 28:50:34.6\\ 
XMMU J050517.3 -- 285023 &05:05:17.3&      -- 28:50:23.7\\ 
XMMU J050522.3 -- 285006 &05:05:22.3&      -- 28:50:06.3\\ 
XMMU J050510.9 -- 284951 &05:05:10.9&      -- 28:49:51.1\\ \hline 
Field RXJ0847.2 + 3449&&\\ 
XMMU J084711.4 + 344717       & 08:47:11.4   & 34:47:17.1\\ 
XMMU J084714.5 + 344654       & 08:47:14.5   & 34:46:54.4\\  
XMMU J084707.5 + 344947       & 08:47:7.5   & 34:49:46.7\\  
XMMU J084709.5 + 344917       & 08:47:9.5   & 34:49:17.1\\ \hline 
Field RXJ1120.1 + 4318&&\\ 
XMMU J111959.1 + 432030  &   11:19:59.1&      43:20:30.3\\ 
XMMU J112001.3 + 431543  &   11:20:01.3   &   43:15:43.2\\ 
XMMU J112004.2 + 431932  &   11:20:04.2  &    43:19:31.6\\ 
XMMU J112008.8 + 432030  &   11:20:08.8  &    43:20:30.3\\ 
XMMU J112009.6 + 432056  &   11:20:09.6  &    43:20:56.4\\ 
XMMU J112014.4 + 431932  &   11:20:14.4  &    43:19:31.6\\ 
XMMU J112015.0 + 432009   &  11:20:15.0   &   43:20:08.5\\ \hline 
Field RXJ1334.3 + 5030&&\\ 
XMMU J133426.3 + 503247  &    13:34:26.3&      50:32:46.8\\ 
XMMU J133410.8 + 503118   &    13:34:10.8&      50:31:17.6\\ 
XMMU J133415.6 + 503030   &   13:34:15.6&      50:30:29.7\\ 
XMMU J133416.3 + 503115     &  13:34:16.3&      50:31:15.4\\ 
XMMU J133430.2 + 503238      & 13:34:30.2&      50:32:38.0\\ 
XMMU J133428.9 + 503141      &13:34:28.9&      50:31:41.5\\ \hline 
Field RXJ1354.2 -- 0222&&\\ 
XMMU J135414.8 -- 022031&13:54:14.8&   -- 02:20:31.7\\ \hline 
\end{tabular} 

\caption{\label{tab:srclist} The identifications (XMM-Newton informal
ID, nominal RA \& Dec ) of point sources that were excised from
the spectral and imaging analysis}
\end{table} 
  
\end{document}